\documentclass{rAMF2e-A}

\usepackage{epstopdf}% To incorporate .eps illustrations using PDFLaTeX, etc.
\usepackage{subfigure}% Support for small, `sub' figures and tables

\begin{document}

%\jvol{00} \jnum{00} \jyear{2015} \jmonth{April}

\articletype{MANUSCRIPT}

\title{Optimising Credit Portfolio Using a Quadratic Nonlinear Projection Method}

\author{
\name{BOGUK KIM$^{\ast}$\thanks{{\em{Correspondence Address}}: Boguk Kim, QMR Group, Inc., Hosu-ro 672, \#1214, Ilsandong-gu, Goyang-si, Gyeonggi-do, 10364, Korea, Republic of. Email: kim.boguk@qmr.re.kr}, CHULWOO HAN$^{\dag}$ \& FRANK C. PARK$^{\ddag}$}
\affil{$^\ast$QMR Group, Inc., Hosu-ro 672, \#1214, Ilsandong-gu, Goyang-si, Gyeonggi-do, 10364, Korea, Republic of., $^{\dag}$Durham University Business School, Mill Hill Lane, Durham, DH1 3LB, UK, $^{\ddag}$School of Mechanical \& Aerospace Engineering, Seoul National University, 1 Gwanak-ro, Gwanak-gu, Seoul 08826, Korea, Republic of.}
\received{received July 2017}
}

\maketitle

\begin{abstract}
A novel optimisation framework through quadratic nonlinear projection is introduced for credit portfolio when the portfolio risk is measured by Conditional Value-at-Risk (CVaR). The whole optimisation procedure to search towards the optimal portfolio state is conducted by a series of single-step optimisations under the local constraints described in the multi-dimensional constraint parameter space as functions of the total amount of portfolio adjustment. Each single-step optimisation is approximated by the first-order variation of the weight increments with respect to the total amount of portfolio adjustment and is solved in the form of locally exact formula formulated in the general Lagrange multiplier method. Our method can deal with optimisation for general nonlinear objective functions, such as the return-to-risk ratio maximisation or the diversification index, as well as the risk minimisation or the return maximisation.
\end{abstract}


\section{Introduction}

The loss distribution of a credit portfolio is generally far from the standard Gaussian. Rather, it is highly non-symmetric and fat-tailed with large skewness and kurtosis. This implies that the mean--variance analysis is not suitable for the credit portfolio optimisation. In such cases, it is appropriate to use tail risk measures, among which Conditional Value-at-Risk (CVaR) is a popular choice as a ``natural coherent alternative to VaR" \citep{RockafellarUryasev2002, AcerbiTasche2001, AcerbiTasche2002, Tasche2002}.

Since CVaR is a convex measure, the associated risk minimisation can be effectively formulated in the form of convex optimisation through the linear programming (LP) \citep{RockafellarUryasev2000, Uryasev2000, KrokhmalPalmquistUryasev2002,  MansiniOgryczakSperanza2007}. The convexity of CVaR guarantees to find the unique global minimum of the portfolio risk, if it exists. It has been reported that there are sheer advantages in minimising portfolio risk employing CVaR over other types of risk measures, according to risk minimisation based on various empirical data in \citet{Pflug2000, GoldbergHayesMahmoud2011}.

Nevertheless, the LP method for the CVaR minimisation has a few critical drawbacks in the following sense:
\begin{itemize}
\item The domain of the risk minimisation, which is the range of possible adjustments of individual assets or asset groups in the portfolio, should be arbitrarily specified in order to include the global extremum point, so the LP method is of no use if there is no global minimum even for the convex risk function. (We refer to an asset group as a collection of assets of similar risk structures in a portfolio; see Figure 1).
\begin{figure}
  \centering
    \includegraphics[width=1\linewidth]{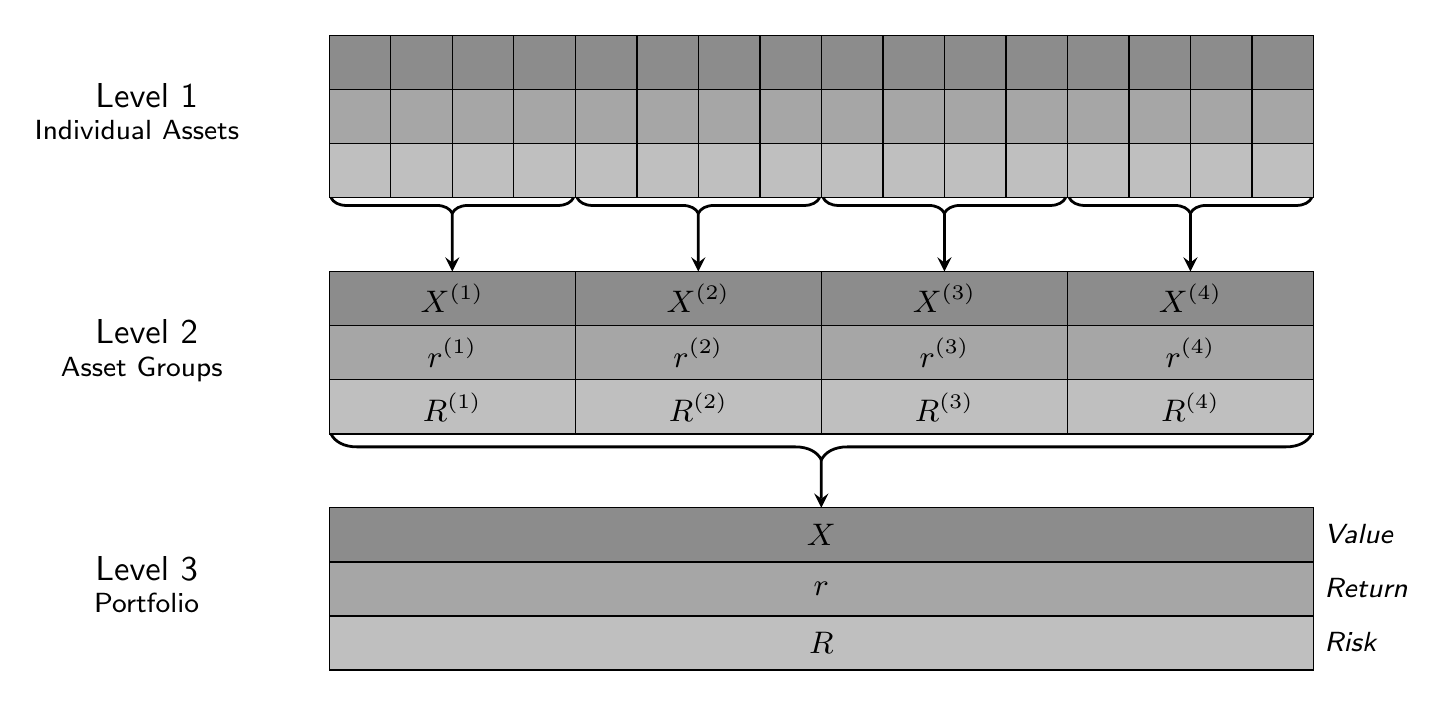}

   \caption{\label{fig:} A schematic diagram for the structure of portfolio, asset groups, individual assets when the number of asset groups is $N=4$.}
\end{figure}

\item There is no known procedure yet to optimise non-convex objective functions, such as the return-to-risk ratio which is among the single most important concepts in the portfolio management, through LP.
\end{itemize}

In general, the portfolio optimisation is essentially a nonlinear problem if the risk measure is any type of Value-at-Risk (VaR), so it is natural to introduce a nonlinear method to handle such a problem. The standard implementation of a nonlinear portfolio optimisation problem, may be cumbersome when there are too many assets or asset groups in a portfolio because the associated Jacobian matrix is usually dense for any portfolio of a fully nontrivial asset correlation. Moreover, it may be necessary to deal with non-smooth or discontinuous loss distributions, for which discontinuous partial derivatives in the Jacobian matrix should be properly treated in the weak sense.

In this paper, we newly introduce an alternative optimisation framework for the credit portfolio optimisation by maximising the return-to-risk index when CVaR is used as the risk measure. The formulation introduced here is expressed via the Lagrange multiplier method contingent upon an artificially introduced small enough quadratic error term, comparable to the total infinitesimal change of asset or asset-group allocation in a portfolio, as the necessary constraint.

The key ingredient is to represent the objective functions with respect to multi-dimensional parameters that correspond to changes in constraints, for instance, the total portfolio revenue and the total portfolio return or the tolerance of additional risk, as functions of the total amount of portfolio adjustment, at the current portfolio state in the weight distribution space, typically higher dimensional for any portfolio optimisation. So to speak, this gives a quadratic map between the first-order variation of the contribution of each asset or asset group in a portfolio and the multiple constraint parameters. Such a mapping process is a natural outcome from a series of local optimisations approximated by the amount of weight distribution changes with respect to the total amount of portfolio adjustment, solved by the Lagrange multiplier method, of which consequence is just a form of quadratic nonlinear projection.

The overall optimisation procedure is conducted by a continuation procedure of these local optimisations with respect to the total amount of portfolio adjustment. Namely, local single-step optimisations are repeatedly performed over and over until a threshold is reached in the total amount of portfolio adjustment (see Figure 2). Each step of the local optimisations can be solved in the closed form formula, owing to which the overall required computation becomes efficient. In that way, our new method enables to maximise or minimise general nonlinear objective function, including the return-to-risk ratio, as a natural extension from the risk minimisation or the return maximisation, also featured by searching the optimal paths under various given constraints even when the global maximum or minimum of the objective function does not exist.

\begin{figure}
  \centering
    \includegraphics[width=1\linewidth]{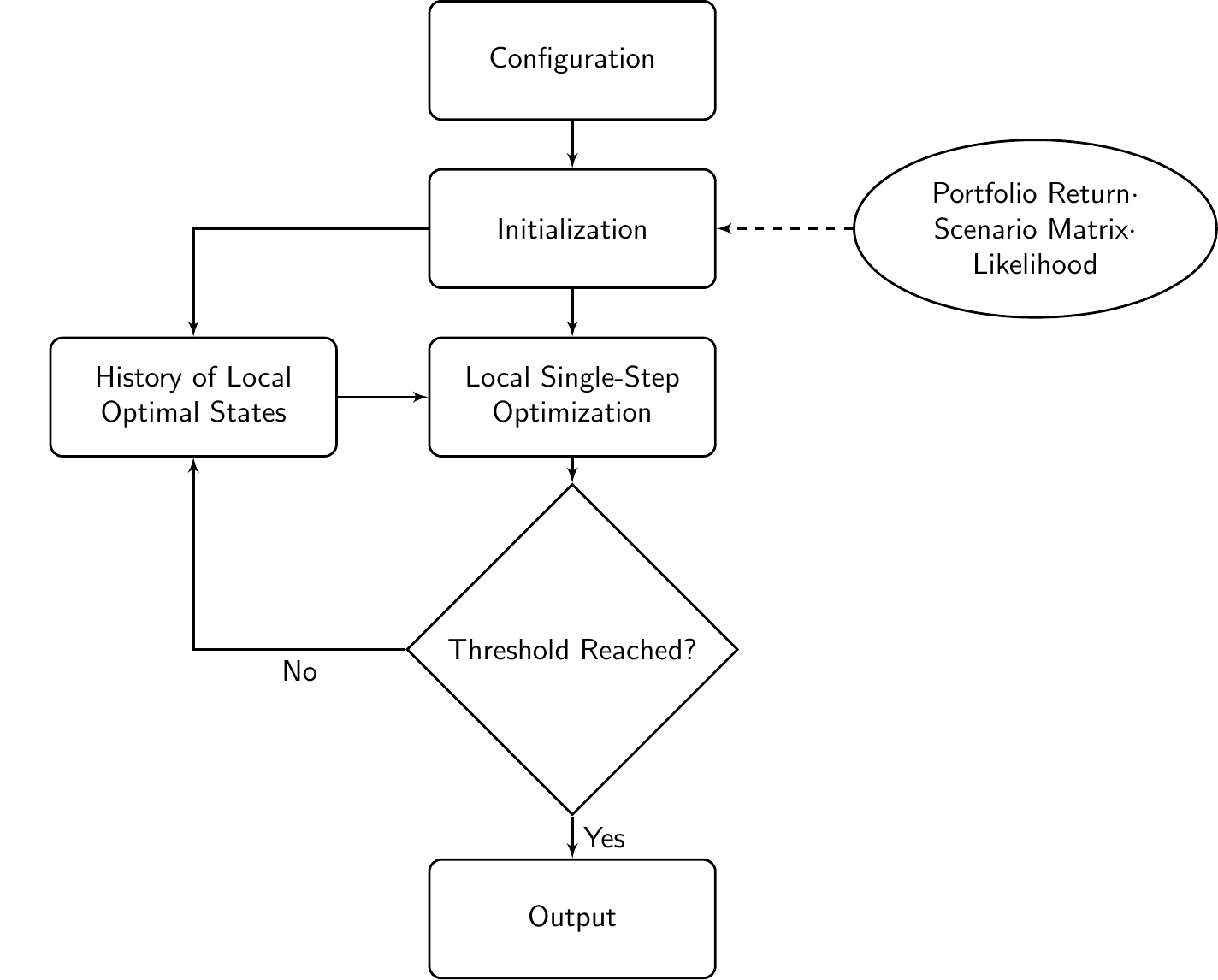}

   \caption{\label{fig:} The main flowchart of the overall portfolio optimisation procedures.}
\end{figure}

\section{Preliminaries}

\subsection{Time series of a portfolio}

Suppose that a credit portfolio consists of $N$ different assets or asset groups for $N\ge 2$. Asset groups are characterised by a collection of assets which are completely or closely related (i.e. the correlation between assets in an asset group is considered to be close to the unity) within themselves (Figure 1). Let us say that a random variable $X$ represents the value of the total portfolio. The value of each asset or asset group is denoted by random variables $X_k^{(n)}$ for $n=1,\,2,\,\cdots,\,N$, where $k$ denotes a parameter representing the time step.

A time series for a virtual scenario of portfolio, starting from the initial portfolio allocation $\left(X_0^{(1)},\, X_0^{(2)},\, \cdots,\, X_0^{(N)}\right)$, can be simply expressed in the following matrix form:
\begin{eqnarray}
\label{scenario_matrix}
\begin{array}{cccc}
X_1^{(1)}&X_1^{(2)}&\cdots&X_1^{(N)}\\
X_2^{(1)}&X_2^{(2)}&\cdots&X_2^{(N)}\\
\vdots&\vdots&\vdots&\vdots\\
X_k^{(1)}&X_k^{(2)}&\cdots&X_k^{(N)}\\
\vdots&\vdots&\vdots&\vdots
\end{array}.
\end{eqnarray}
$X_k=\sum_{n=1}^{N}X_k^{(n)}$ is defined as the total portfolio value at the $k$-th step. If the time series is generated for $K$ steps, where $K$ is a natural number, from a pre-specified joint  probability distribution, then the occurrence likelihood at the $k$-th step is given by $e_k$ for $k=1,\,2,\,\cdots,K$ such that $\sum_{k=1}^{K}e_k=1$. A finite or possibly the infinite number of rows correspond to time steps within a time horizon, for instance, 6 months, 1 year, and 2 years, etc., when each value of asset or asset group is estimated. Without the loss of generality, we may assume that the $N$ columns are linearly independent, namely, there are no assets or asset groups of our interest that are expressed by linear superpositions of the other assets or asset groups.

\subsection{Return}

Let us say that the return of each asset or asset group is denoted by $r^{(n)}$ for $n=1,\,2,\,\cdots,\,N$. Then, the relation between individual returns and the total return is simply given by
\begin{eqnarray}
r&=&\sum\limits_{n=1}^Nr^{(n)}w^{(n)},
\end{eqnarray}
where $w^{(n)}=\frac{X^{(n)}}{X_0}$ is the initial weight of the $n$-th asset or asset group for $n=1,\,2,\,\cdots,\,N$ with respect to the total portfolio. The return of each asset or asset group may be actually adjusted in the process of portfolio reallocation.

\subsection{Probability distribution of loss}

The loss distribution of a portfolio is obtained by counting the frequencies---possibly weighted with the occurrence likelihood pre-specified---of the loss in the time series, where the loss of the $n$-th individual asset or asset group is defined by $Z^{(n)}_k=X_0^{(n)}-X^{(n)}_k$ for $n=1,\,2,\,\cdots,\,N$ and $k=1,\,2,\,\cdots$ (accordingly, $Z_k=X_0-X_k$ for the whole portfolio).
The correlations between different assets or asset groups are naturally deduced from the set of times series. The key assumption is that a unique convergent loss distribution function exists, along with the associated correlation coefficients that are fixed, regardless of the proportion of each asset or asset group in a portfolio.

\subsection{Risk measures}

\subsubsection{Value-at-risk (VaR)}

VaR of the total portfolio for its loss $Z=X_0-X$ is defined in terms of its loss distribution function ${\mathrm P}(\cdot)$ by
\begin{eqnarray}
{\rm VaR}_{\beta}^{(\rm loss)}(X)\equiv {\rm VaR}_{\beta}(Z)=\inf_{Y\in \mathcal{A}}\left\{{\rm P}\left(\left.Z\right|Z\le Y\right)\ge\beta\right\}
\end{eqnarray}
for a given confidence level $0\le \beta\le 1$, where $\mathcal{A}$ is the admissible set of the portfolio loss, which is the set of real numbers. Typical values for the confidence level $\beta$ used in real practice are 0.95, 0.99, 0.995, 0.999, etc. within a given time horizon. There are also many other risk measures associated with VaR.

\subsubsection{Conditional value-at-risk (CVaR) for loss}

CVaR of the total portfolio is defined in terms of the underlying VaR by
\begin{eqnarray}
{\rm CVaR}_{\beta}^{({\rm loss})}(X)=\frac{{\rm CVaR}_{\beta}^{(\rm loss)+}(X)+(\beta^*-\beta){\rm VaR}_{\beta}^{({\rm loss})}(X)}{1-\beta},
\end{eqnarray}
where
\begin{subequations}
\begin{eqnarray}
\beta^*&=&{\rm P}\left(\left.Z\right|Z< {\rm VaR}_{\beta}^{(\rm loss)}(X)\right),\\
{\rm CVaR}_{\beta}^{(\rm loss)+}(X)&=&{\rm E}\left[\left.Z\right| Z\ge{\rm VaR}_{\beta}^{({\rm loss})}\left(X\right)\right]\equiv\int^{+\infty}_{{\rm VaR}_{\beta}^{({\rm loss})}\left(X\right)}Z{\rm P}(Z){\rm d}Z.
\end{eqnarray}
\end{subequations}
Note that CVaR$^{+}_{\beta}\left(\cdot\right)$ is called the tail conditional expectation. Either when the loss distribution is continuous or when $\beta$ does not split any of atoms in the discrete loss distribution, in particular, we have $\beta^*=\beta$, so that ${\rm CVaR}^{(\rm loss)}_{\beta}\left(\cdot\right)={\rm CVaR}^{({\rm loss})+}_{\beta}\left(\cdot\right)$. Obviously, CVaR$^{(\rm loss)}_{\beta}\left(\cdot\right)$ is no less than VaR$^{(\rm loss)}_{\beta}\left(\cdot\right)$.

\subsubsection{Risk contribution and Derivative-at-Risk (DaR)}

For the purpose of portfolio optimisation, it is useful to consider the risk contributions of individual assets or asset groups with respect to the underlying risk measure of the total portfolio. This is feasible if the total risk is measured from a time series of assets or asset groups in a portfolio.

When the total risk is measured by CVaR, the risk contribution of the $n$-th asset or asset group is given by
\begin{eqnarray}
{\rm CVaR}^{({\rm loss})^{(n)}}_{\beta}(X)&=&{\rm E}\left[\left. Z^{(n)} \right| Z={\rm CVaR}_{\beta}^{({\rm loss})}\left(X\right)\right],
\end{eqnarray}
where $Z^{(n)}=X^{(n)}_0-X^{(n)}$ is the loss of the $n$-th asset or asset group for $n=1,\,2,\,\cdots,\,N$. Therefore, we should have
\begin{eqnarray}
\sum\limits_{n=1}^{N}{\rm CVaR}^{({\rm loss})^{(n)}}_{\beta}(X)={\rm E}\left[\left.Z\right| Z={\rm CVaR}^{({\rm loss})}_{\beta}\left(X\right)\right]={\rm CVaR}^{({\rm loss})}_{\beta}\left(X\right).
\end{eqnarray}

The underlying tail risk measure is simple scale-invariant, namely, ${\rm CVaR}_{\beta}^{(\rm loss)}\left((1+\delta a)X\right)=(1+\delta a){\rm CVaR}_{\beta}^{(\rm loss)}(X)$, where $|\delta \alpha|\ll 1$. Hence, it follows that
\begin{eqnarray}
{\rm CVaR}^{({\rm loss})^{(n)}}_{\beta}(X)=X^{(n)}\frac{\partial {\rm CVaR}_{\beta}^{(\rm loss)}(X)}{\partial X^{(n)}}=w^{(n)}\frac{\partial {\rm CVaR}_{\beta}^{(\rm loss)}(X)}{\partial w^{(n)}},
\end{eqnarray}
where the partial derivatives, named as Derivative-at-Risk (DaR), exist \citep{Tasche2000}. In such a case, $\frac{\partial {\rm CVaR}_{\beta}^{(\rm loss)}(X)}{\partial w^{(n)}}$ exists in the distribution sense, in other words, in the sense of weak derivative.

Because CVaR is piecewise linear with respect to the proportion of individual assets or asset groups in a portfolio, its associated partial derivatives are all piecewise constant functions. Accordingly, the relation between CVaR and DaR is derived as follows:
\begin{eqnarray}
{\rm CVaR}_{\beta}^{({\rm loss})}(X)=\sum\limits_{n=1}^{N}w^{(n)}{{\rm DaR}^{(\rm loss)}}^{(n)}_{\beta}(X),
\end{eqnarray}
where ${{\rm DaR}^{(\rm loss)}}^{(n)}_{\beta}(X)=\frac{\partial {\rm CVaR}_{\beta}^{({\rm loss})}(X)}{\partial w^{(n)}}$.

\subsection{Return-to-risk index}

It is customary that we may expect more return from more risk. In order to measure the performance level of assets, asset groups, and the whole portfolio, it is desirable to adopt a performance measure that is defined by the ratio between the return and the risk as follows:
\begin{subequations}
\begin{eqnarray}
\label{RORAC_total}
\mathcal{I}&\equiv&\frac{rX_0}{{\rm CVaR}^{({\rm loss})}_{\beta}\left(X\right)},\\
\label{RORAC_partial}
\mathcal{I}^{(n)}&\equiv&\frac{r^{(n)}X_0^{(n)}}{{\rm CVaR}^{({\rm loss})}_{\beta}\left(X^{(n)}\right)}\quad{\rm for}\quad n=1,\,2,\,\cdots,\,N
\end{eqnarray}
\end{subequations}
for the total portfolio and each individual asset or asset group, respectively. This index is naturally a nonlinear function of the weight of individual asset or asset group for a general risk measure, but it becomes a rational function made of a ratio between two piecewise linear functions with respect to the weight functions for CVaR.

\subsection{Diversification index}

The ratio of the total portfolio risk with respect to the sum of the risk of assets or asset groups is defined as the diversification index ${\mathcal D}_{\beta}(X)$:
\begin{eqnarray}
{\mathcal D}_{\beta}\left(X\right)=\frac{{\rm CVaR}^{({\rm loss})}_{\beta}\left(X\right)}{\sum\limits_{n=1}^N{\rm CVaR}^{({\rm loss})}_{\beta}\left(X^{(n)}\right)}=\frac{\sum\limits_{n=1}^Nw^{(n)}{{\rm DaR}_{\beta}^{(\rm loss)}}^{(n)}\left(X\right)}{\sum\limits_{n=1}^N{\rm CVaR}^{({\rm loss})}_{\beta}\left(X^{(n)}\right)}.
\end{eqnarray}
This quantity measures the degree of correlation between the losses of individual assets or asset groups. As a special case, if the assets or asset groups are completely correlated, then they all behave like a single asset such that ${\rm DaR}^{({\rm loss})^{(n)}}_{\beta}(X)={\rm CVaR}^{({\rm loss})}_{\beta}\left(X^{(n)}\right)$ for $n=1,\,2,\,\cdots,\,N$, so that ${\mathcal D}_{\beta}(X)$ should be 1, but this critical case is excluded in our discussion. The convexity of CVaR ensures that ${\mathcal D}_{\beta}(X)$ is a positive number, nontrivially less than 1.

\section{Quadratic optimisation framework}

The fundamental question is how to adjust the initial asset allocation in order to maximise or to minimise a given objective function. More specifically, what should be the most fair way to enhance the performance of a portfolio from the current asset allocation for a given amount of the total portfolio adjustment?

Let us begin our main discussion by denoting $w_{+}^{(n)}$ as the adjusted weight of the $n$-th asset of asset group in a portfolio after an infinitesimal amount of the asset or asset group adjustment process, $w_{-}^{(n)}$ as the original weight of the $n$-th asset of asset group in a portfolio, and $\delta w^{(n)}_{\pm}=w_{+}^{(n)}-w_{-}^{(n)}$ for $n=1,\,2,\,\cdots,\,N$. Also, we define that $\vec{w}_{\pm}=\left(w_{\pm}^{(1)},w_{\pm}^{(2)},\cdots,w_{\pm}^{(N)}\right)^{\rm T}$ and $\delta\vec{w}_{\pm}=\left(\delta w_{\pm}^{(1)},\delta w_{\pm}^{(2)},\cdots,\delta w_{\pm}^{(N)}\right)^{\rm T}$ for the whole portfolio: $X^{(n)}_{\pm}$, $X_{\pm}$,
$r_{\pm}$, $\delta r_{\pm}$, ${\mathcal I}^{(n)}_{\pm}$, $\delta \mathcal{I}^{(n)}_{\pm}$, $\mathcal{I}_{\pm}$, $\delta \mathcal{I}_{\pm}$ for $n=1,\,2,\,\cdots,\,N$ are defined all in the same fashion. It is additionally assumed that each weight component is nonzero, namely $w_{\pm}^{(n)}\ne 0$ for $n=1,\,2,\,\cdots,\,N$, because there is no need to consider reallocating the assets or asset groups of no contribution for any practical purpose.

\subsection{Objective functions}
We introduce the following objective functions for the single-step optimisation procedure in the forward sense:
\begin{itemize}
\item{Risk minimisation (min Ri)}

\begin{eqnarray}
\label{Sum_DaR_to_w}
\min_{\delta{\vec{w}}}\frac{\delta {\rm CVaR}^{({\rm loss})}_{\beta}\left(X_{\pm}\right)}{X_0},
\end{eqnarray}
where
\begin{eqnarray}
\delta {\rm CVaR}^{({\rm loss})}_{\beta}\left(X_{\pm}\right)=\sum_{n=1}^{N}{{\rm DaR}_{\beta}^{(\rm loss)}}^{(n)}\left(X_{-}\right)\delta w^{(n)}_{\pm},
\end{eqnarray}
assuming that $\delta w^{(n)}$ is small enough so that the partial derivative $\frac{\partial {\rm CVaR}^{({\rm loss})}_{\beta}\left(X_{\pm}\right)}{\partial w^{(n)}}$ stays unchanged before and after the adjustment of an infinitesimal asset allocation. Note that dividing by $X_0$ is just for the purpose of nondimensionalsation.

\item{Return maximisation (max Re)}

\begin{eqnarray}
\max_{\delta{\vec{w}}}\sum\limits_{n=1}^{N}r^{(n)}_{-}\delta w^{(n)}_{\pm}.
\end{eqnarray}

\item{Return-to-risk index maximisation (max Re2Ri)}

\begin{eqnarray}
\label{Alternative_objective_function}
\max_{\delta{\vec{w}}}\delta\left\{\frac{r_{\pm}X_0}{{\rm CVaR}^{({\rm loss})}_{\beta}\left(X_{\pm}\right)}\right\}.
\end{eqnarray}

\item{Diversification index minimisation (min DI)}

\begin{eqnarray}
\label{Alternative_objective_function}
\min_{\delta{\vec{w}}}\delta\left\{\frac{\sum\limits_{n=1}^Nw^{(n)}{{\rm DaR}_{\beta}^{(\rm loss)}}^{(n)}\left(X_{\pm}\right)}{\sum\limits_{n=1}^N{\rm CVaR}^{({\rm loss})}_{\beta}\left(X^{(n)}_{\pm}\right)}\right\}.
\end{eqnarray}
\end{itemize}

\subsection{Constraints}
Depending on the objective function, we may select the following constraints in the optimisation procedure:
\begin{itemize}%[$\bullet$]
\item Constraint on the total revenue:
\begin{eqnarray}
\sum\limits_{n=1}^{N}\delta w^{(n)}_{\pm}=\delta\alpha.
\end{eqnarray}

\item Constraint on the total return (except for max Re):
\begin{eqnarray}
\sum\limits_{n=1}^{N}r^{(n)}_{-}\delta w^{(n)}_{\pm}=\delta\gamma,
\end{eqnarray}
where
\begin{eqnarray}
\sum\limits_{n=1}^{N}r^{(n)}_{\pm}w^{(n)}_{\pm}=r_{\pm}.
\end{eqnarray}

\item Constraint on the total risk (except for min Ri):
\begin{eqnarray}
\frac{1}{X_0}\sum\limits_{n=1}^{N}{{\rm DaR}_{\beta}^{(\rm loss)}}^{(n)}\left(X_{-}\right)\delta w^{(n)}_{\pm}=\delta\gamma.
\end{eqnarray}

\item Constraint on the (weighted) amount of weight adjustment in the $l_2$ sense:
\begin{eqnarray}
\sqrt{\sum\limits_{n=1}^{N}{c^{(n)}_{-}}^2\delta {w^{(n)}_{\pm}}^2}= \delta c,
\end{eqnarray}
where $c^{(n)}$ is the coefficient of amount of weight adjustment of each asset or asset group for $n=1,\,2,\,\cdots,\,N$.
\end{itemize}
Indeed, we may choose the return-to-risk index or the diversification index as the constraints, as well, as long as they are not taken as the objective functions.

We may define the cost constraint in the $l_p$ sense for any $p> 0$, but $p=2$ is chosen, according to which the optimisation method is referred to be as a quadratic nonlinear projection method.

In this method, the amount of weight adjustment constraint must be taken for any types of optimisation  whereas the other types of constraints may be excluded. For the optimisation of the return-to-risk indices, either constraint on the total risk and/or on the total return can be specified. The coefficients of amount of weight adjustment may be subject to further modelling.

\subsection{Local single-step optimisation of the first-order variational approximation under the constraints of infinitesimal asset adjustment}

For our optimisation procedure, we consider the change of the first-order variation of the objective function with respect to the infinitesimal total amount of portfolio adjustment, $\delta c$, which is a measure for the total cost of weight adjustment in the $l_2$ sense for $n=1,\,2,\,\cdots,\,N$. The constraint on the total amount of portfolio adjustment is always required in our optimisation framework. For the sake of brevity of description, we constrict our discussion to the cases when the total revenue, the total risk, or the total return are selected for additional constraints. Here, we refer to $\kappa_1$ and $\kappa_2$ as the path parameters for the the total revenue adjustment ratio and the total return or risk adjustment ratio with respect to $\delta c$, respectively.

The whole optimisation procedure is completed by a continual sequence of single-step optimisations, with respect to $\delta c$ in the forward sense, each of which is formulated through the Lagrange multiplier method.

To this end, we assume that the initial weight of each asset or asset group is nonzero, namely, $w_0^{(n)}\ne 0$ for $n=1,\,2,\,\cdots,\,N$. Supposing also that $r^{(n)}_{\pm}=r^{(n)}$ and $c^{(n)}_{\pm}=c^{(n)}$ for $n=1,\,2\,\cdots,\,N$ are all unchanged between before and after any single-step optimisation for the sake of simplicity, the common framework of the associated Lagrange multiplier method is given as follows when the cost adjustment constraint is defined in $l_2$ norm:
\begin{eqnarray}
\label{LagrangeMultiplierMethod_Main}
\mathcal{L}\!=\!\!\sum\limits_{n=1}^{N}f^{(n)}y^{(n)}\!\!-\!\!\left(\sum\limits_{n=1}^{N}y^{(n)}\!-\!\kappa_1\!\!\right)\!s\!-\!\left(\sum\limits_{n=1}^{N}h^{(n)}y^{(n)}\!-\!\kappa_2\!\right)\!t\!-\!\left(\!\!\sqrt{\sum\limits_{n=1}^{N}{c^{(n)}}^2{y^{(n)}}^2}\!-\!1\!\right)\!q,
\end{eqnarray}
for
\begin{subequations}
\label{LagrangeMultiplierMethod_VariableParameters}
\begin{eqnarray}
y^{(n)}&=&\frac{\delta w^{(n)}_{\pm}}{\delta c},\\
\kappa_1&=&\frac{\delta\alpha}{\delta c},\\
\kappa_2&=&\frac{\delta\gamma}{\delta c},
\end{eqnarray}
\end{subequations}
where
\begin{subequations}
\label{LagrangeMultiplierMethod_Eqns}
\begin{eqnarray}
\frac{\delta \mathcal{L}}{\delta y^{(n)}}&=&0\quad {\rm for}\quad n=1,\,2,\,\cdots,\,N,\\
\frac{\partial \mathcal{L}}{\partial s}&=&0,\\
\frac{\partial \mathcal{L}}{\partial t}&=&0,\\
\label{L_Eq_p}
\frac{\partial \mathcal{L}}{\partial q}&=&0.
\end{eqnarray}
\end{subequations}

Then, the single-step optimisation is solved as follows (see Appendix A):
\begin{enumerate}[$\bullet$]
\begin{subequations}
\item when both of the total revenue and the total risk or return constraints are included:
\begin{eqnarray}
\label{expression_dw_dc}
y^{(n)}\!=\!\frac{f^{(n)}}{q{c^{(n)}}^2}\!-\!\frac{GW\!-\!HV\!+\!\left(HU\!-\!GV\right)h^{(n)}}{(UW\!-\!V^2)q{c^{(n)}}^2}\!-\!\frac{\kappa_2 V\!-\!\kappa_1 W\!+\!(\kappa_1 V\!-\!\kappa_2 U)h^{(n)}}{(UW\!-\!V^2){c^{(n)}}^2}
\end{eqnarray}
for
\begin{eqnarray}
q&=&\pm\sqrt{-\frac{a_0}{a_2}}
\end{eqnarray}
with
\begin{eqnarray}
\label{expression_a0}
a_0&=&F-\frac{H^2U+G^2W-2GHV}{UW-V^2},\\
\label{expression_a2}
a_2&=&\frac{U\kappa_2^2+W\kappa_1^2-2V\kappa_1\kappa_2}{UW-V^2}-1,
\end{eqnarray}
\item when the total risk or return constraint is excluded:
\begin{eqnarray}
\label{expression_dw_dc_option2}
y^{(n)}&=&\frac{f^{(n)}}{q{c^{(n)}}^2}-\frac{G}{Uq{c^{(n)}}^2}+\frac{\kappa_1}{U{c^{(n)}}^2},\\
q&=&\pm\sqrt{-\frac{a_0}{a_2}}
\end{eqnarray}
for
\begin{eqnarray}
\label{expression_a0_option2}
a_0&=&F-\frac{G^2}{U},\\
\label{expression_a2_option2}
a_2&=&\frac{\kappa_1^2}{U}-1,
\end{eqnarray}
\item when the total revenue constraint is excluded:
\begin{eqnarray}
\label{expression_dw_dc_option3}
y^{(n)}&=&\frac{f^{(n)}}{q{c^{(n)}}^2}-\frac{Hh^{(n)}}{Wq{c^{(n)}}^2}+\frac{\kappa_2h^{(n)}}{W{c^{(n)}}^2},\\
q&=&\pm\sqrt{-\frac{a_0}{a_2}}
\end{eqnarray}
for
\begin{eqnarray}
\label{expression_a0_option3}
a_0&=&F-\frac{H^2}{W},\\
\label{expression_a2_option3}
a_2&=&\frac{\kappa_2^2}{W}-1,
\end{eqnarray}
\item when both of the total revenue and the total risk or return constraints are excluded:
\begin{eqnarray}
\label{expression_dw_dc_option4}
y^{(n)}&=&\pm\frac{f^{(n)}}{\sqrt{F}{c^{(n)}}^2}
\end{eqnarray}
\end{subequations}
\end{enumerate}
for $n=1,\,2,\,\cdots,\,N$, where
\begin{subequations}
\begin{eqnarray}
\label{Constant_F}
F&=&\sum\limits_{n=1}^{N}\frac{{f^{(n)}}^2}{{c^{(n)}}^2},\\
\label{Constant_G}
G&=&\sum\limits_{n=1}^{N}\frac{f^{(n)}}{{c^{(n)}}^2},\\
\label{Constant_H}
H&=&\sum\limits_{n=1}^{N}\frac{f^{(n)}h^{(n)}}{{c^{(n)}}^2},\\
\label{Constant_U}
U&=&\sum\limits_{n=1}^N\frac{1}{{c^{(n)}}^2},\\
\label{Constant_V}
V&=&\sum\limits_{n=1}^N\frac{h^{(n)}}{{c^{(n)}}^2},\\
\label{Constant_W}
W&=&\sum\limits_{n=1}^N\frac{{h^{(n)}}^2}{{c^{(n)}}^2}.
\end{eqnarray}
\end{subequations}
$a_0$ is always positive for non-trivial portfolios, thus, it is required that $a_2<0$ in order to make $q$ real-valued, so that any possible degenerate cases, where there are infinitely many solutions, are excluded. $\delta c$ should be always nonzero to proceed the optimisation process when the portfolio is in any sub-optimal states even though the other two path parameters $\kappa_1$ and $\kappa_2$ are allowed to be zero. In summary, the local weight change $\delta\vec{w}_{\pm}$ at each single-step optimisation is given by a function of $\left(\kappa_1,\kappa_2\right)$ (see Figure 3). In case that each $w^{(n)}$ should stay to be non-negative for $n=1,\,2,\,\cdots,\,N$, we stop the update procedure only for those of weight components that hit the zero value.

\begin{figure}
  \centering
 \includegraphics[width=0.4\linewidth]{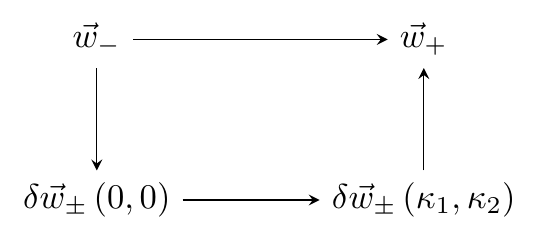}

\caption{\label{fig:} The update procedure for the local single-step optimisation through the quadratic nonlinear projection. The local weight change $\delta\vec{w}_{\pm}$, or equivalently $\frac{\delta\vec{w}_{\pm}}{\delta c}$, is a function of $\left(\kappa_1,\kappa_2\right)$, and the current state $\vec{w}_-$ corresponds to when $\left(\kappa_1,\kappa_2\right)=(0,0)$. For an appropriate pair $(\kappa_1,\kappa_2)$, $\delta\vec{w}_{\pm}$ is determined by solving the associated quadratic equation, and then $\vec{w}_{+}=\vec{w}_{-}+\delta\vec{w}_{\pm}\left(\kappa_1,\kappa_2\right)$ is imposed.}
\end{figure}

Once we have two real solutions for $q$, we choose only one of them so that it satisfies the condition of the objective function: we take the larger value of the objective function to find the desired local optimal state for the return maximisation and the smaller one for the risk minimisation (see Figure 4).

\begin{figure}

\centering{
 \includegraphics[width=1\linewidth]{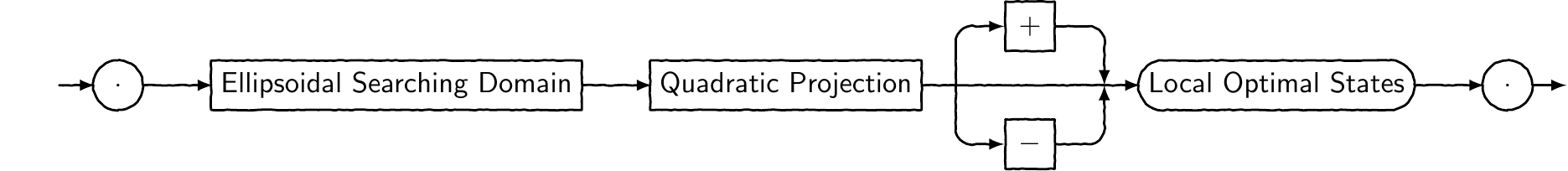}}

   \caption{\label{fig:} The single-step optimisation procedures via the quadratic nonlinear projection. The local optimal states are achieved by following the only one of the two paths, denoted by $+$ and $-$, which correspond to the two different real solutions of the quadratic equations.}
\end{figure}

In particular, when $\kappa_1 = 0$, this quadratic optimisation problem is understood as finding the maximum and the minimum, which are unique respectively on the crosscut between an $(N-1)$-dimensional hyper-plane and the surface of an $N$-dimensional ellipsoid centred at the origin (see Figure 5).

\begin{figure}
  \centering
 \includegraphics[width=1\linewidth]{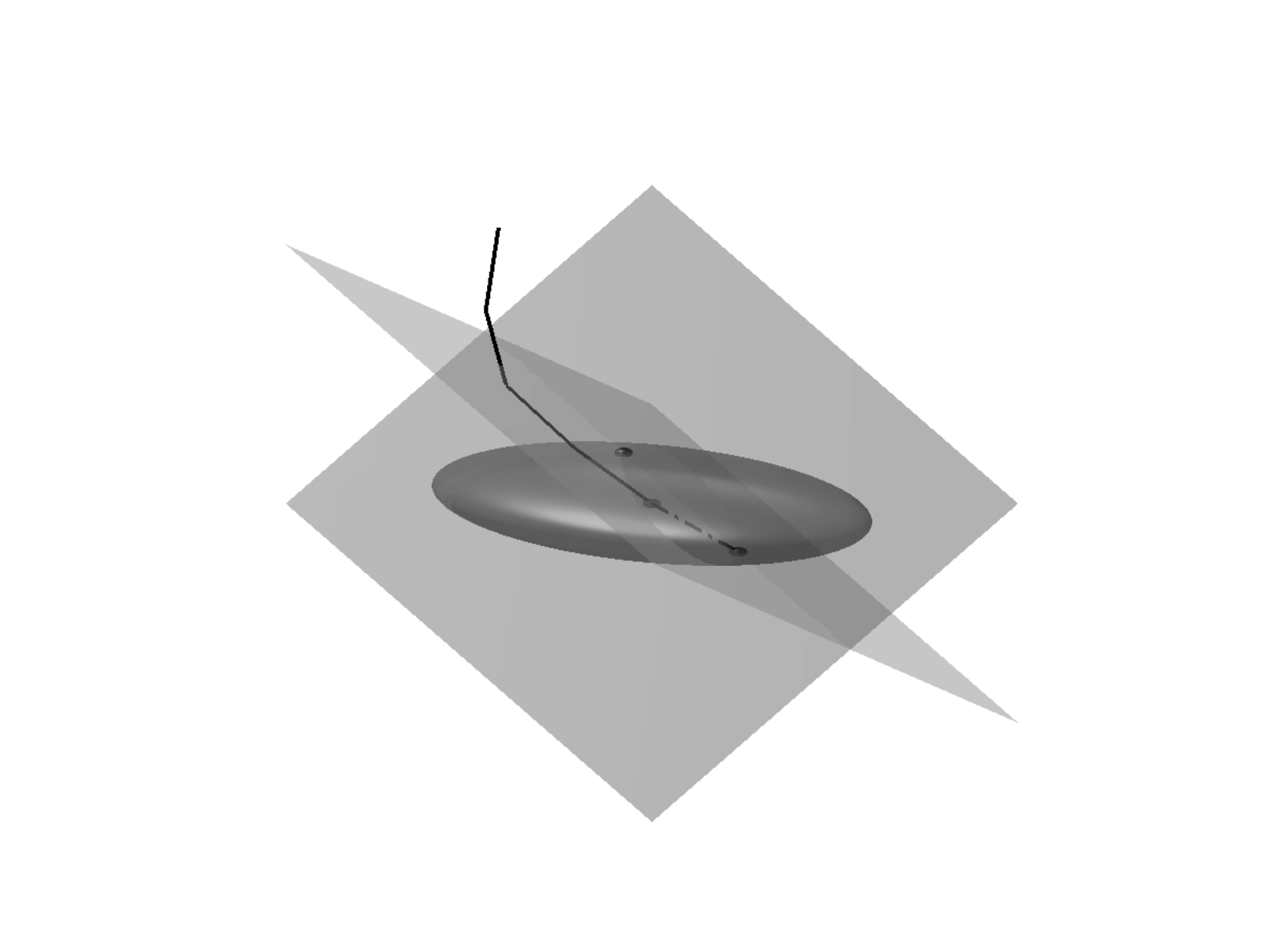}

   \caption{\label{fig:} A diagram for local single-step optimisation via the quadratic nonlinear projection when $N=3$. The surface of ellipsoid: the range of local searching for optimal states. The center of ellipsoid: the current portfolio state. Two hyper-planes: the set of two constraints. The two intersections of the surface of ellipsoid and two hyper-planes: the two optimal states as the solutions to the Lagrange multiplier method via the quadratic equation. The new extension (-$\cdot$-) of discrete approximation of the optimal path (---). }
\end{figure}

Depending on the objective functions and constraints, $f^{(n)}$, $h^{(n)}$ for $n=1,\,2,\,\cdots,\,N$ can be chosen as follows:
\begin{subequations}
\begin{eqnarray}
f^{(n)}&=&\left\{\!\!\!
\begin{array}{cl}
r^{(n)} &\!\!\!\!\! \mbox{for max Re},\\
\frac{{{\rm DaR}_{\beta}^{(\rm loss)}}^{(n)}\left(X_{-}\right)}{X_0} &\!\!\!\!\! \mbox{for min Ri},\\
\left(r^{(n)}-\frac{r_-{{\rm DaR}_{\beta}^{(\rm loss)}}^{(n)}\left(X_{-}\right)}{{\rm CVaR}_{\beta}^{(\rm loss)}(X_-)}\right)\frac{X_0}{{\rm CVaR}_{\beta}^{(\rm loss)}(X_-)}&\!\!\!\!\! \mbox{for max Re2Ri},\\
\frac{{\rm DaR}^{({\rm loss})^{(n)}}_{\beta}\left(X_{-}\right)}{\sum\limits_{n=1}^N{\rm CVaR}^{({\rm loss})}_{\beta}\left(X_-^{(n)}\right)}-\frac{{\rm CVaR}^{({\rm loss})}_{\beta}\left(X_0^{(n)}\right){\rm CVaR}^{({\rm loss})}_{\beta}\left(X_-\right)}{\left\{\sum\limits_{n=1}^N{\rm CVaR}^{({\rm loss})}_{\beta}\left(X_-^{(n)}\right)\right\}^2} &\!\!\!\!\! \mbox{for min DI},
\end{array}
\right.\\
h^{(n)}&=&\left\{\!\!\!
\begin{array}{cl}
\frac{{{\rm DaR}_{\beta}^{(\rm loss)}}^{(n)}\left(X_{-}\right)}{X_0}& \!\!\!\!\!\mbox{for max Re},\\
r^{(n)} & \!\!\!\!\!\mbox{for min Ri or min DI}.
\end{array}
\right.
\end{eqnarray}
\end{subequations}
Note that the information of $h^{(n)}$ for $n=1,\,2,\,\cdots,\,N$ is not required for the return-to-risk maximisation because it is essentially identical to the return maximisation if the risk constraint is pre-specified or to the risk minimisation if the return constraint is pre-assigned.

\subsection{Extension to numerical continuation}

The quadratic optimisation is conducted by a sequence of aforementioned approximated optimisation for infinitesimal amount of weight adjustments. For given constraints, all parameters are divided into multiple subdivisions that make a discretised approximation associated with a numerical continuation path, represented by the following sequences:
\begin{eqnarray}
\left\{\left.\left(\alpha_{m},\gamma_{m}, c_{m}\right)\right\vert\left.\alpha_{0}=0,\;\gamma_{0}=0,\;0=c_{0}<c_{m}<c_{M}\right.\right\}_{m=0}^{M},
\end{eqnarray}
and
\begin{subequations}
\begin{eqnarray}
\left(\delta\alpha_{m},\delta\gamma_{m},\delta c_{m}\right)&=&\left(\alpha_{m}-\alpha_{m-1},\gamma_{m}-\gamma_{m-1},c_{m}-c_{m-1}\right),\\
\left(\kappa_{1,m},\kappa_{2,m}\right)&=&\frac{\left(\delta\alpha_{m},\delta\gamma_{m}\right)}{\delta c_{m}}
\end{eqnarray}
\end{subequations}
for $m=1,\,2,\,\cdots,\,M$. Each $\delta c_{m}$ for $m=1,\,2,\,\cdots,\,M$ should be small enough for each single-step optimisation, so that it can satisfy $a_2<0$.

At the end of each single-step optimisation between subintervals of the continuation path, the total risk of the whole portfolio, as well as the risk contribution of each asset or asset group, should be updated by re-sorting the time series of the adjusted total portfolio. When the fixed total risk constraint is used, in particular, we scale the local optimal weight $\vec{w}_+$ by $\vec{w}\cdot\frac{{\rm CVaR}_{\beta}^{\rm (loss)}\left(X_-\right)}{{\rm CVaR}_{\beta}^{\rm (loss)}\left(X_+\right)}$ and then re-evaluate $r_+$, $X_+$, etc, in order that the total portfolio risk should be unchanged during the optimisation procedure. When the iterative procedures go through all subdivisions in the pre-specified cost parameters, the whole numerical continuation process is completed.

\section{Optimal paths}

The increment ratio of the objective function with respect to the total amount of portfolio adjustment, denoted by $Q=\sum_{n=1}^{N}f^{(n)}y^{(n)}$, is calculated to be
\begin{itemize}
\begin{subequations}
\item when both of the total revenue and the total risk or return constraints are included:
\begin{eqnarray}
\label{expression_Q}
Q&=&\frac{a_0}{q}+\frac{(GW-HV)\kappa_1+(HU-GV)\kappa_2}{UW-V^2},
\end{eqnarray}
\item when the risk or return constraint is excluded:
\begin{eqnarray}
\label{expression_Q_option2}
Q&=&\frac{a_0}{q}+\kappa_1,
\end{eqnarray}
\item when the total revenue constraint is excluded:
\begin{eqnarray}
\label{expression_Q_option3}
Q&=&\frac{a_0}{q}+\kappa_2,
\end{eqnarray}
\item when both of the total revenue and the total risk or return constraints are excluded:
\begin{eqnarray}
\label{expression_Q_option3}
Q&=&\frac{F}{q},
\end{eqnarray}
\end{subequations}
\end{itemize}
where $a_0$ and $q$ are all differently defined, as in Section 3.2, depending on the choice of constraints.

Accordingly, it is apparent that $Q$ is a function of $(\kappa_1,\kappa_2)$, so we calculate the extremum points $(\bar{\kappa}_1,\bar{\kappa}_2)$ of $Q$, at which $Q$ is maximised or minimised, in order to find the parameter curves maximising or minimising the objective functions. The extremum points for the optimal paths are provided in Appendix B.

\section{Numerical results from actual credit portfolio data}

We apply this quadratic nonlinear projection method for proprietary credit portfolio data from a bank in South Korea. The bank's credit portfolio risk data are produced by CreditMetrics. The loss distribution for the credit portfolio is provided during one year time horizon. For our test problems, we use $N=252$ asset groups and $K=2000$ scenarios for $M=10^4$ iterations with $\delta c=10^{-5}$. All coefficients of amount of weight adjustment are assumed to be normalised to the unity.

In Figure 6, the CVaR minimisation results are compared by following optimal parameter paths under different conditions of constraints. For the cases of fixed total revenue and fixed total return, the total risk of portfolio reaches a steady state, at which the total risk attains a local minimum and the total return a local maximum, so those curves sharply flatten, as the total amount of portfolio adjustment grows (see Table 1). Moreover, the risk minimisation naturally implies the diversification of the portfolio risk.

\begin{figure}
  \centering
 \includegraphics[width=0.475\linewidth]{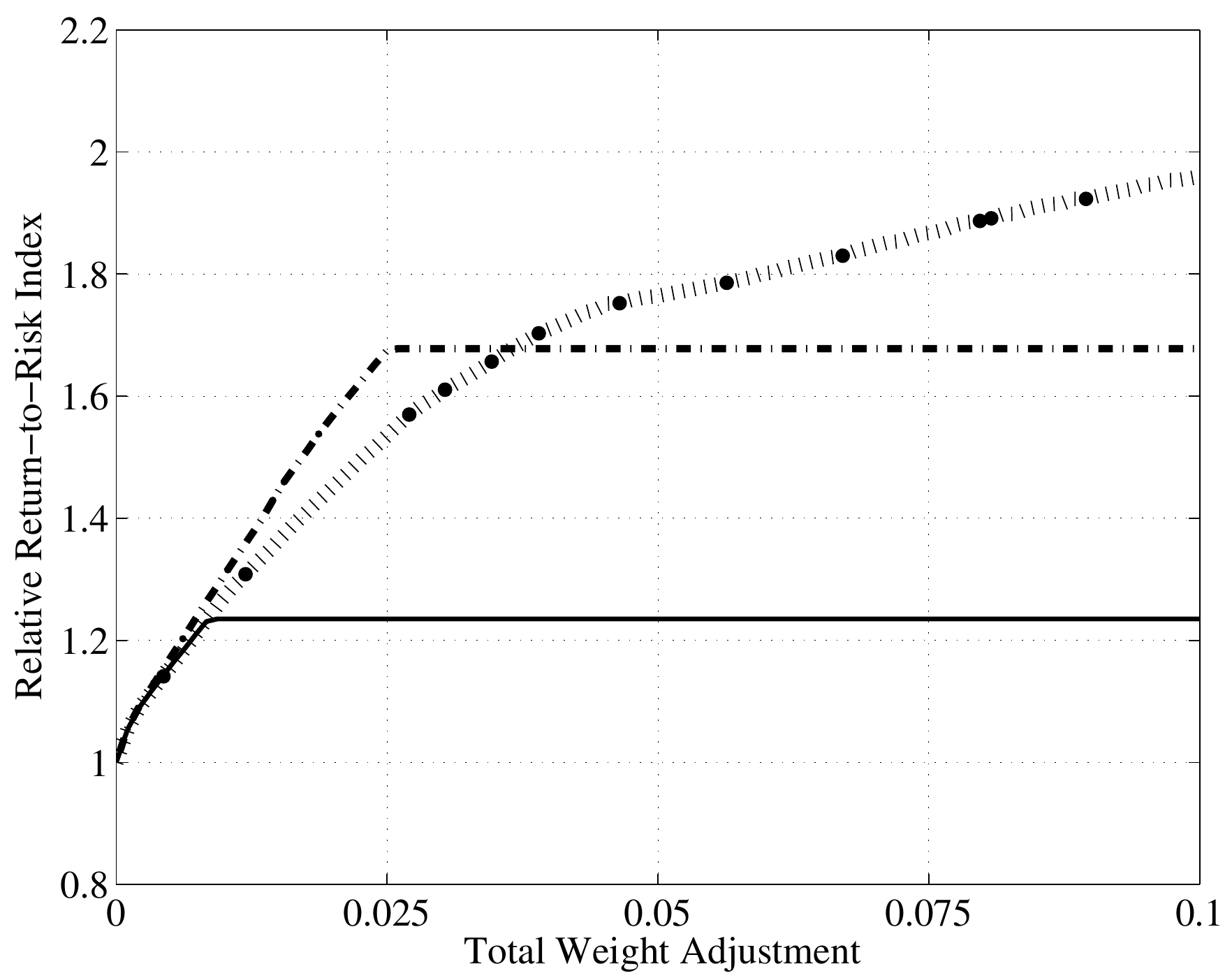} \quad
 \includegraphics[width=0.475\linewidth]{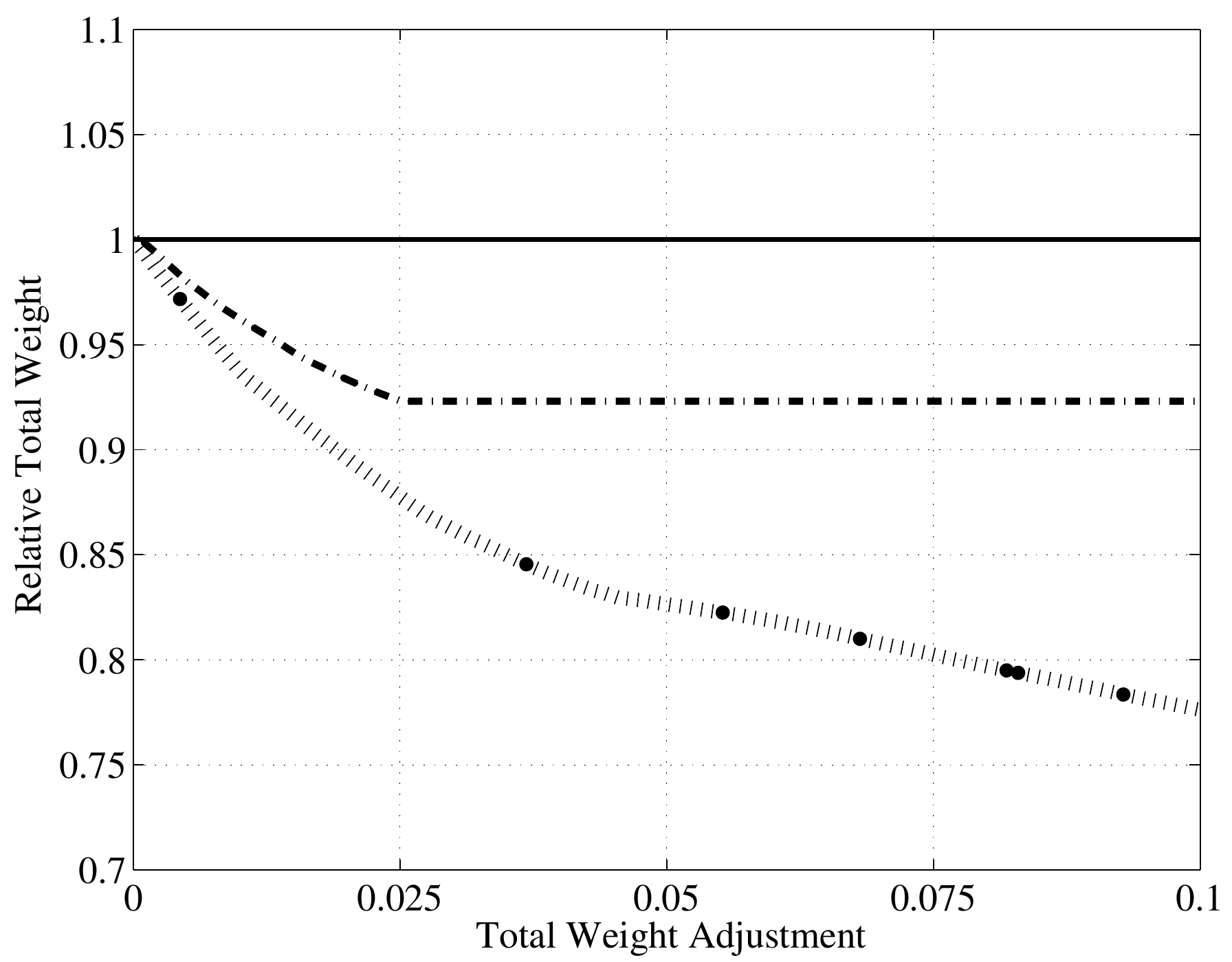} \\
 \includegraphics[width=0.475\linewidth]{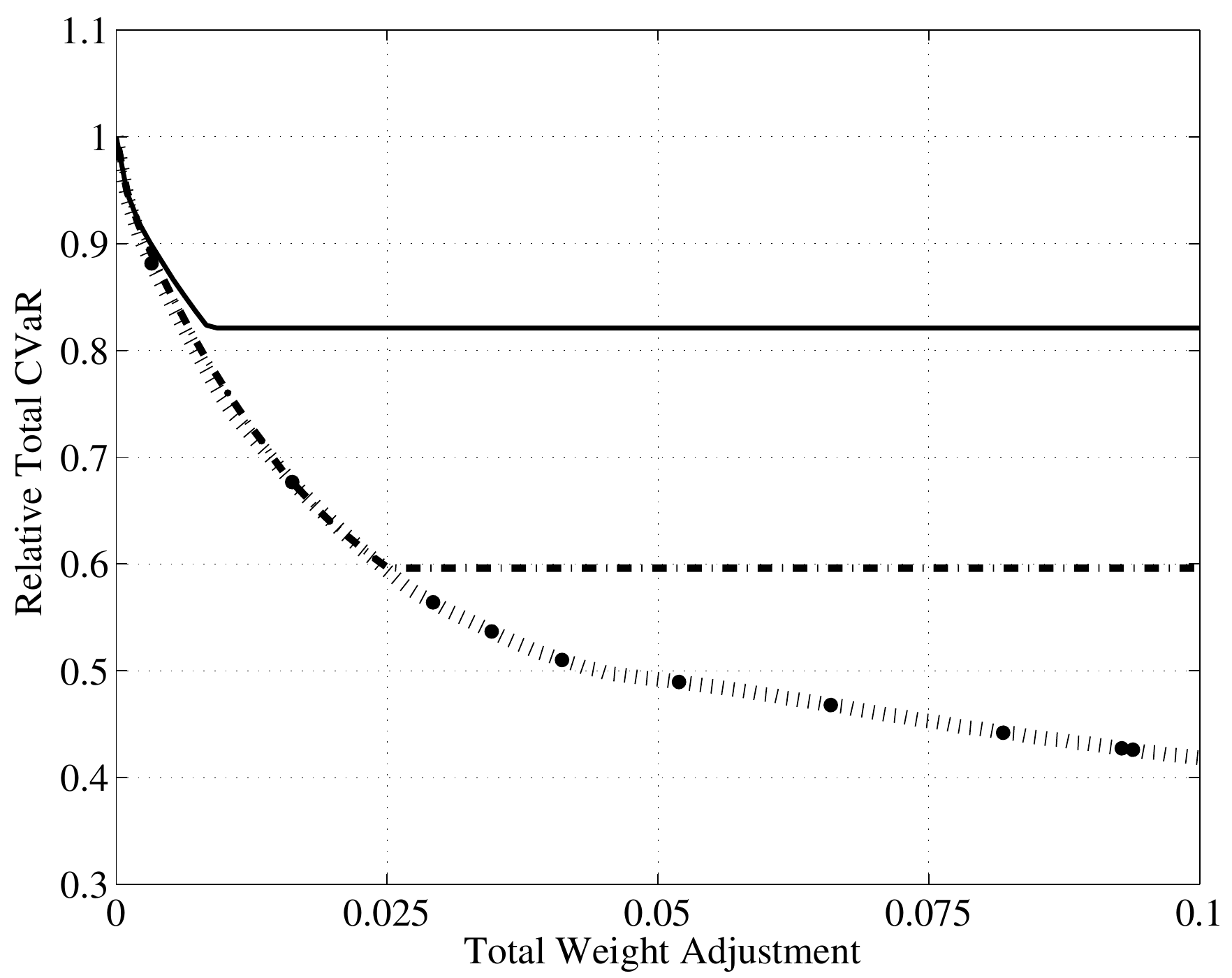} \quad
 \includegraphics[width=0.475\linewidth]{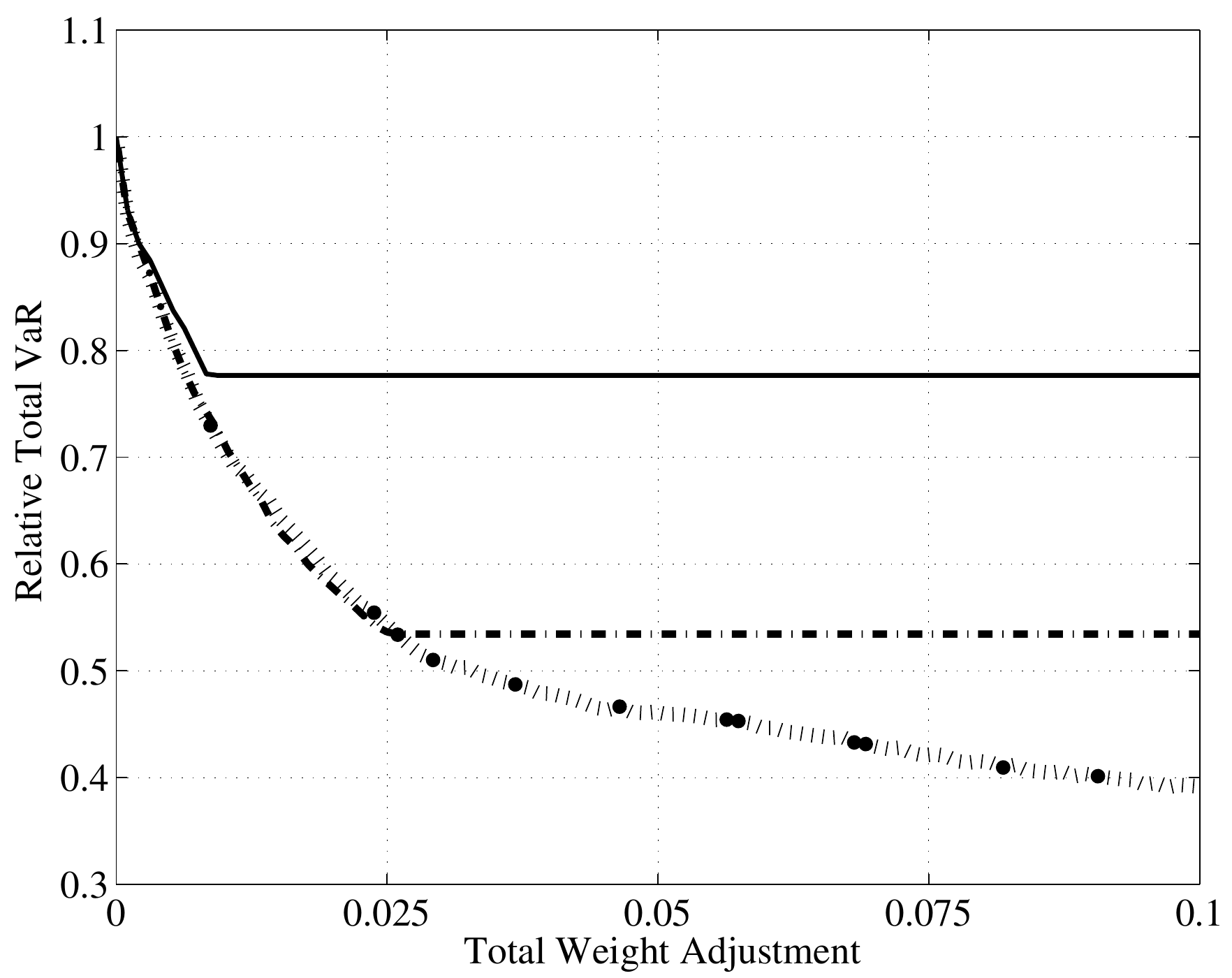} \\
  \includegraphics[width=0.475\linewidth]{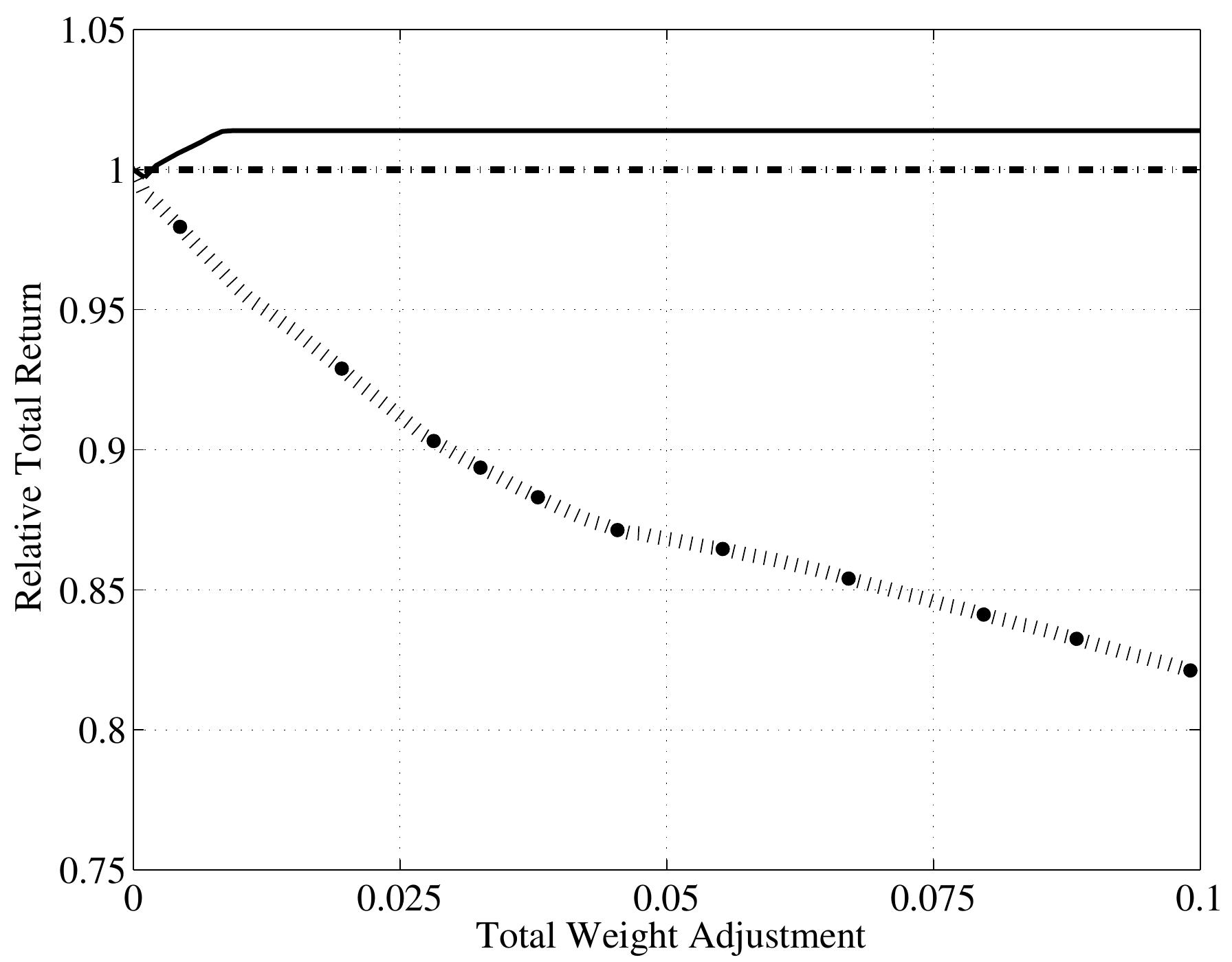} \quad
 \includegraphics[width=0.475\linewidth]{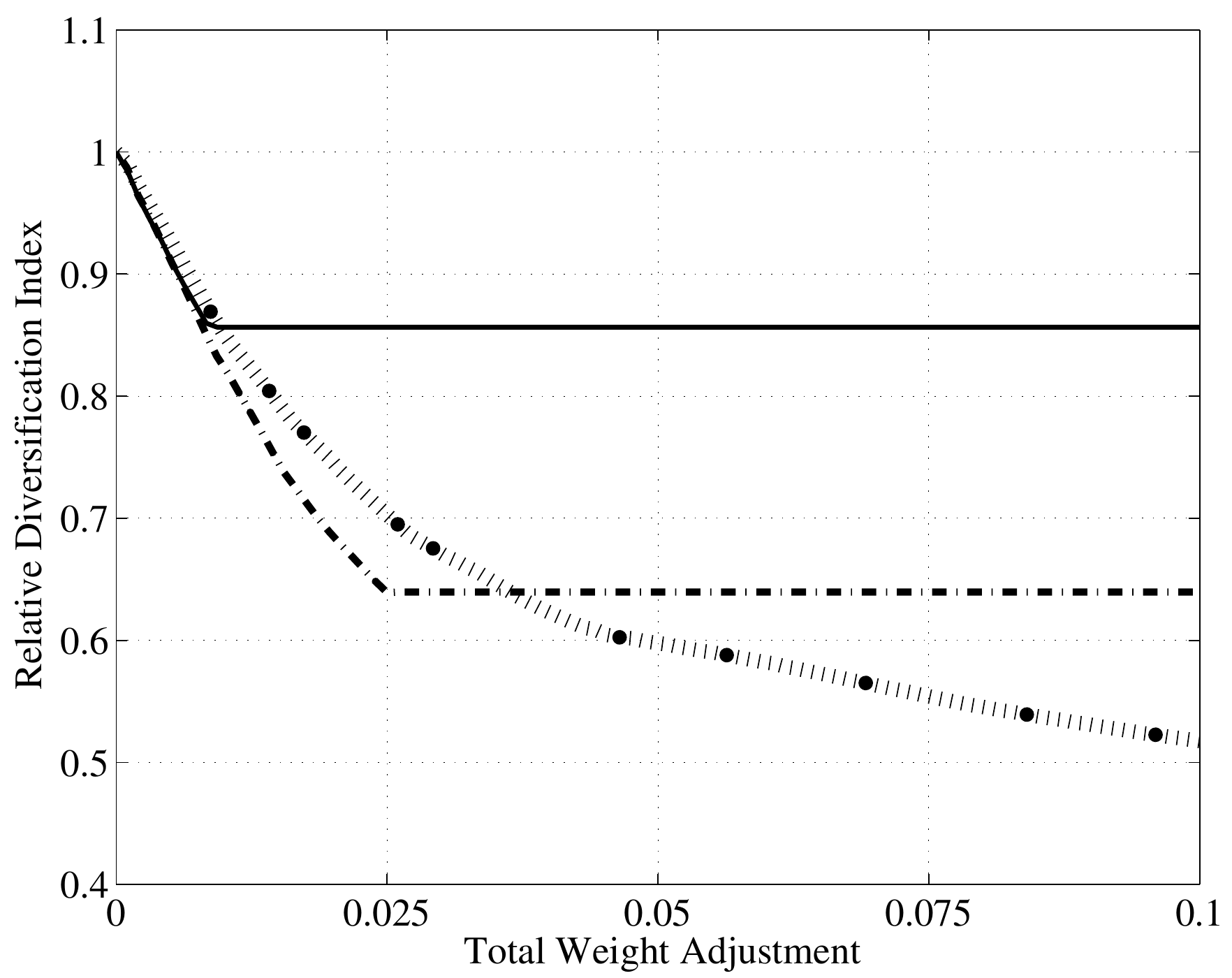} \\
   \caption{\label{fig:} Comparison of risk minimisation results relative with respect to the initial states under different constraints: fixed total revenue (---), fixed total return (-$\cdot$-), and no constraints ($\cdots$) by varying the total amount of portfolio adjustment from 0 to 0.1 when $\delta c=10^{-5}$.}
\end{figure}

\begin{table}
\tbl{The changes of the total risk of our sample portfolio, depending on the objective functions, when the total amount of portfolio adjustment grows from 0 to 0.1 when $\delta c=10^{-5}$.}
{\begin{tabular}[l]{@{}lccc}\toprule
Objective Function &  $c_{M}=0$ & $c_{M}=0.05$ & $c_{M}=0.1$\\
\colrule
min Ri (fixed total revenue) & $2.31402\times 10^{12}$ & $1.89960\times 10^{12}$& $1.89960\times 10^{12}$\\
min Ri (fixed total return) & $2.31402\times 10^{12}$ & $1.37919\times 10^{12}$& $1.37919\times 10^{12}$\\
max Re2Ri & $2.31402\times 10^{12}$ & $1.23949\times 10^{12}$& $1.15843\times 10^{12}$\\
\botrule
\end{tabular}}
\label{table1}
\end{table}

Figure 7 presents the comparisons of the return maximisation results under the constraints of fixed total revenue, fixed total risk, and non-fixed total revenue and non-fixed total risk. Contrary to the risk minimisation, the total risk grows linearly with respect to the total amount of portfolio adjustment, unless the total risk is fixed, as the total portfolio return is being maximised. The return maximisation loses diversification, in other words, increasing the relative diversification index. For the diversification index curve for the return maximisation under the fixed total risk, there are some wiggles, supposedly within the range of numerical approximation error. Under the constraint of the fixed total portfolio risk, the total portfolio revenue should be much reduced so that the total portfolio return becomes also reduced, as well, even for the return maximisation.

\begin{figure}
  \centering
  \includegraphics[width=0.475\linewidth]{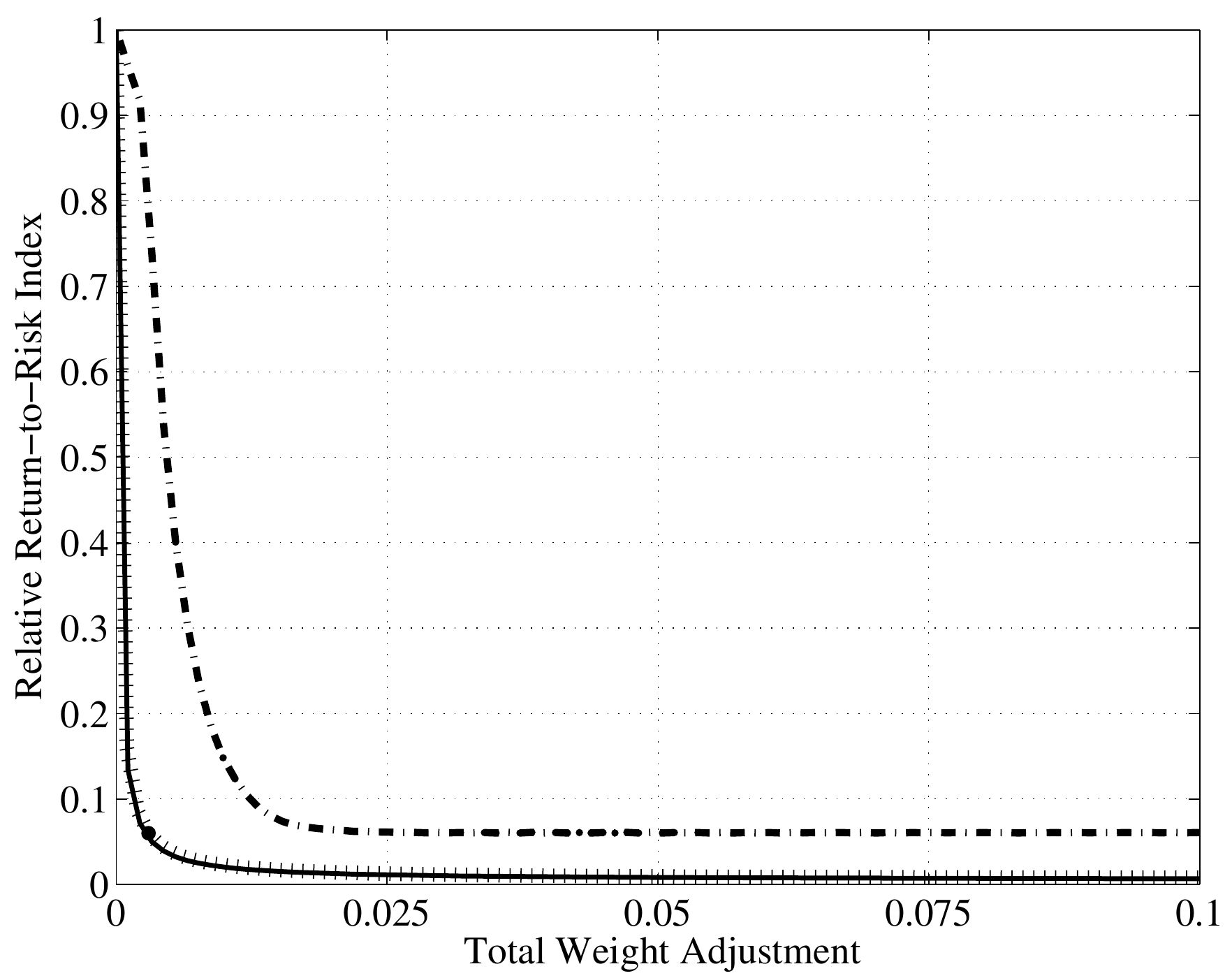}\quad
  \includegraphics[width=0.475\linewidth]{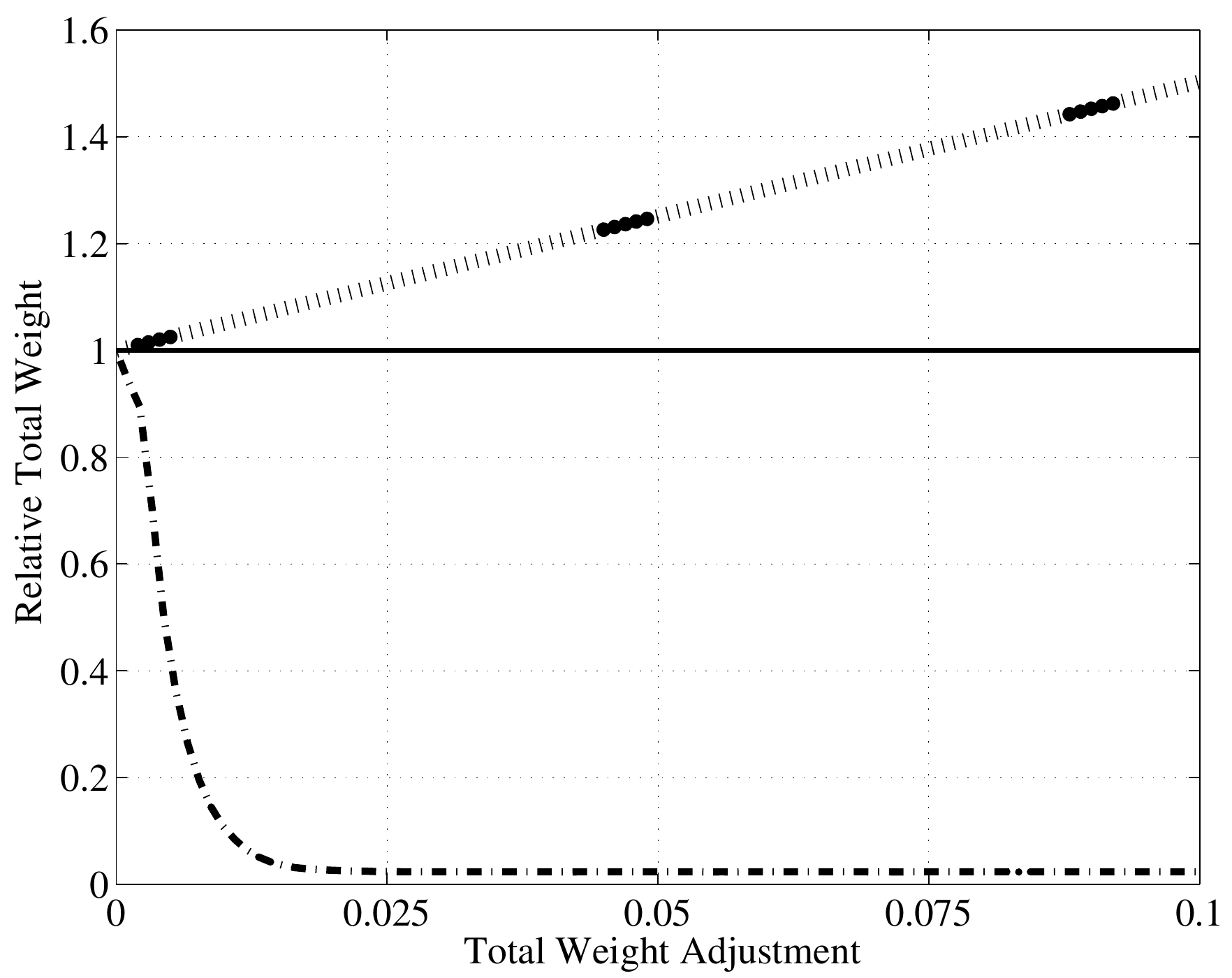}\quad\\
  \includegraphics[width=0.475\linewidth]{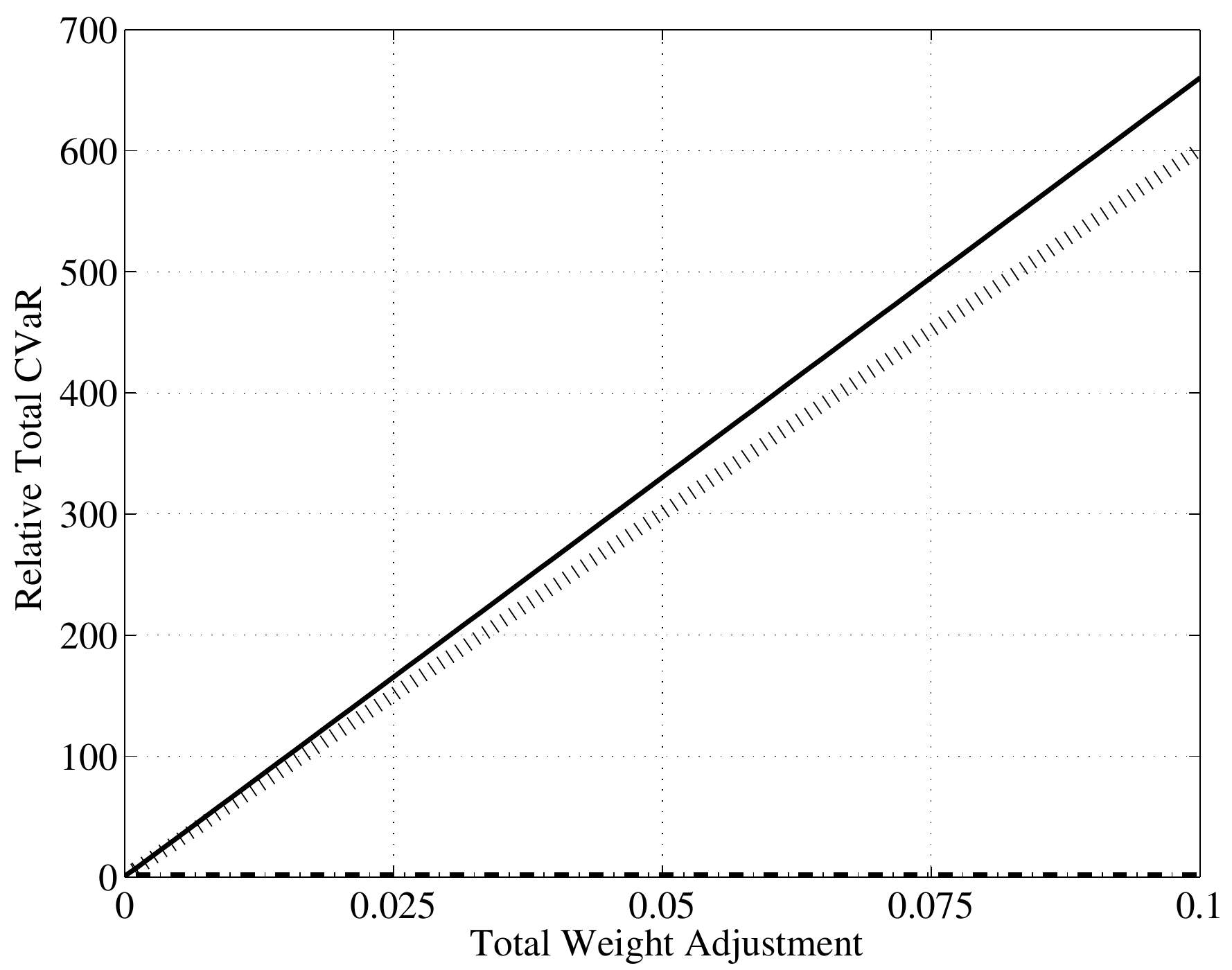}\quad
  \includegraphics[width=0.475\linewidth]{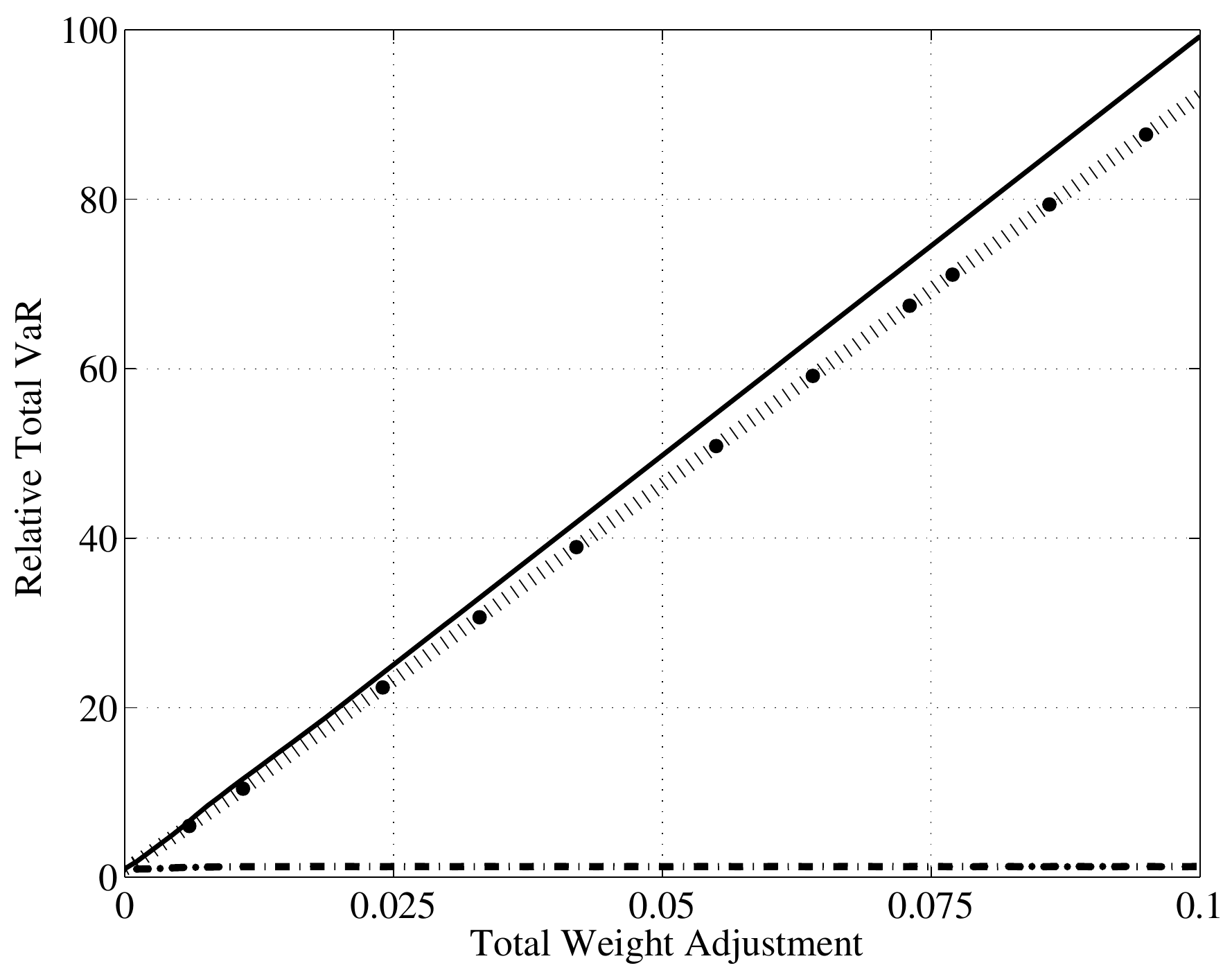}\quad\\
  \includegraphics[width=0.475\linewidth]{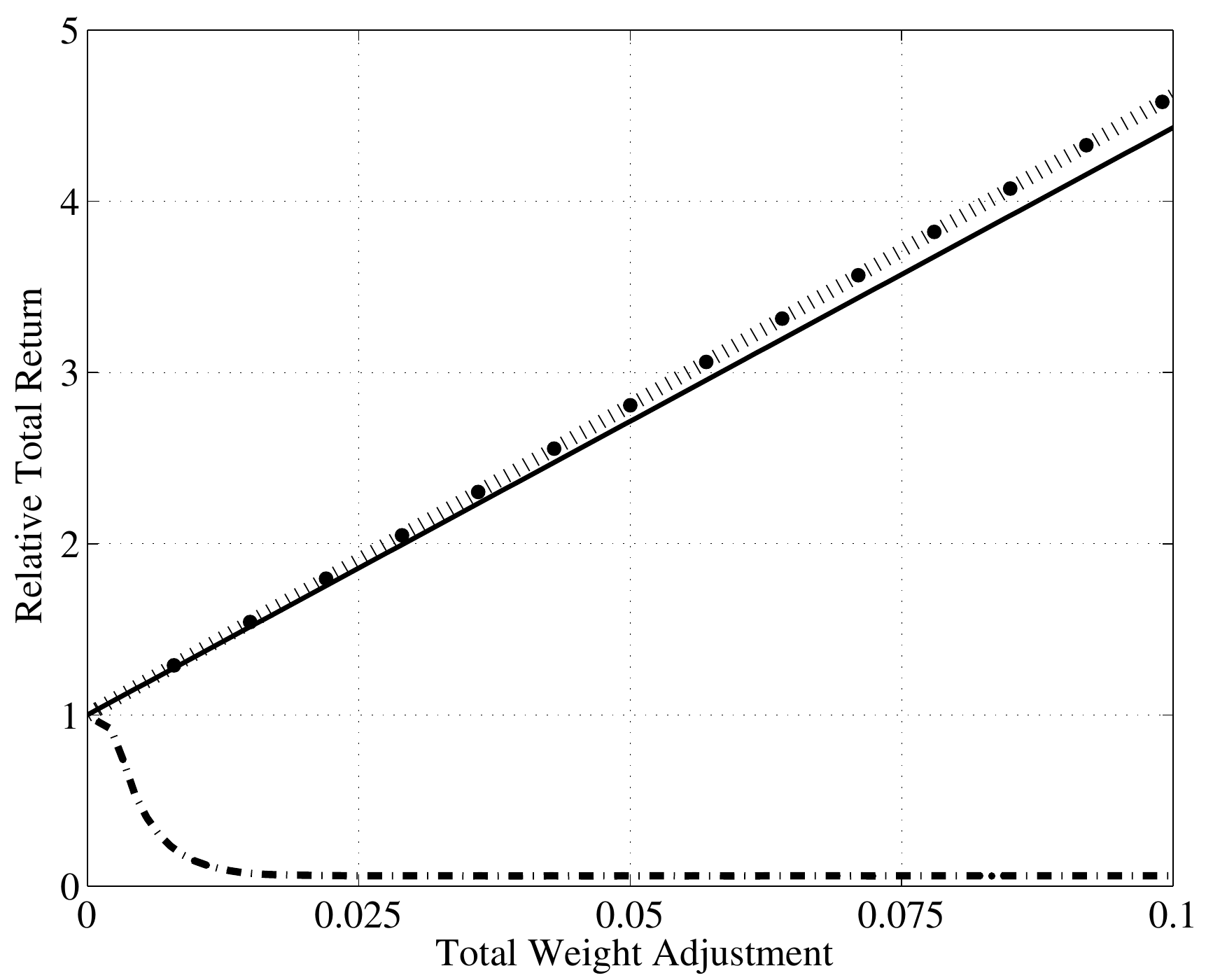}\quad
  \includegraphics[width=0.475\linewidth]{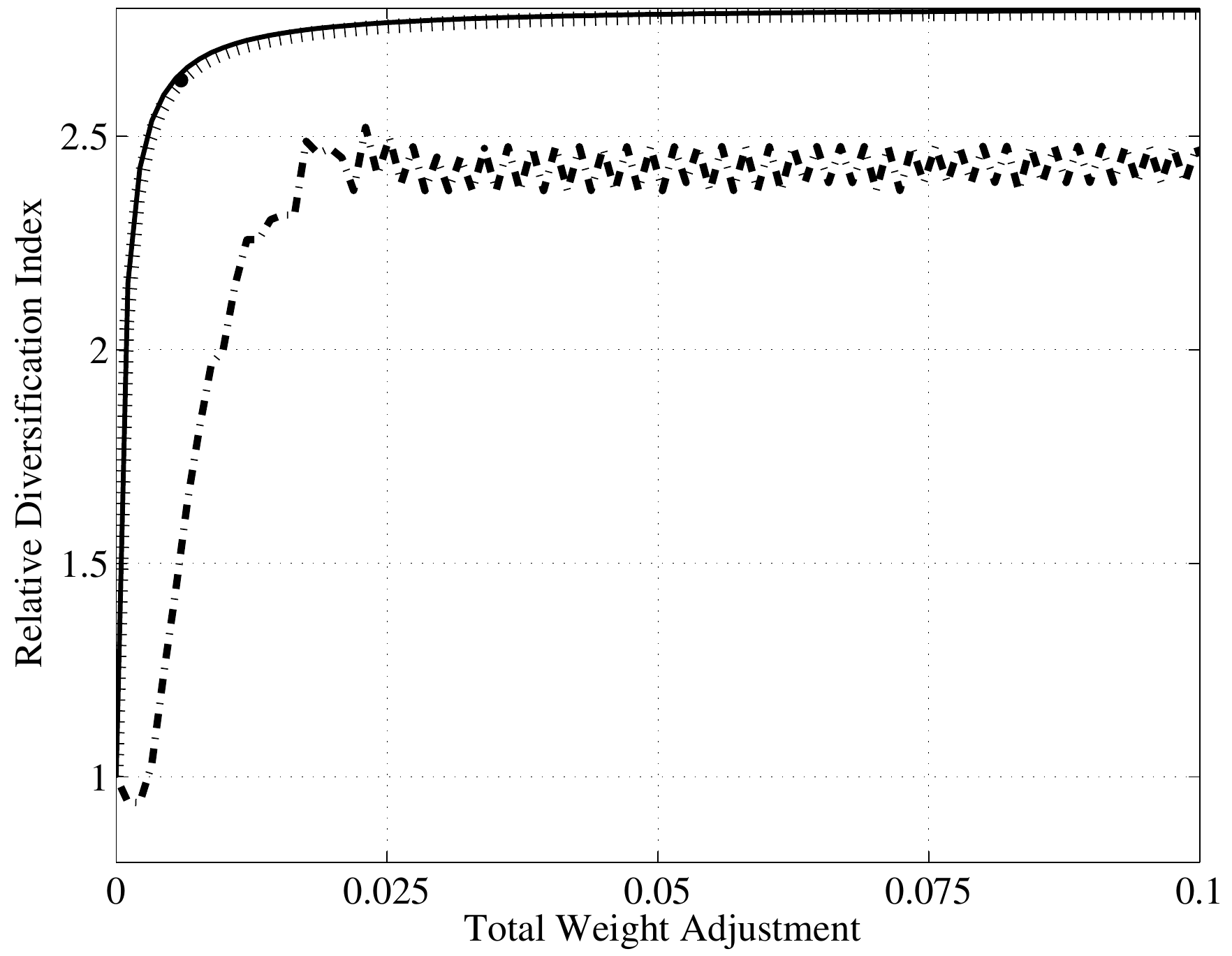}\quad\\
   \caption{\label{fig:} Comparison of return maximisation results relative with respect to the initial states under different constraints: fixed total revenue (---), fixed total risk (-$\cdot$-), and no constraints ($\cdots$) by varying the total amount of portfolio adjustment from 0 to 0.1 when $\delta c=10^{-5}$.}
\end{figure}

In Figure 8, the return-to-risk maximisation results are presented for two different constraints. When the total portfolio revenue is fixed, the return-to-risk maximisation curve is supposedly the efficient frontier, which is the ultimate upper bound of all possible such curves starting from the same initial portfolio state, within the range of numerical error bound. A surprising result from the return-to-risk maximisation is that the total portfolio risk can be further reduced than the result from the risk minimisation for the fixed total revenue condition. This means that increasing the total portfolio return with a slight amount may enhance the total portfolio risk much more than otherwise, depending on the scenario matrix, sometimes even by reducing the total portfolio revenue.

\begin{figure}
 \centering
  \includegraphics[width=0.475\linewidth]{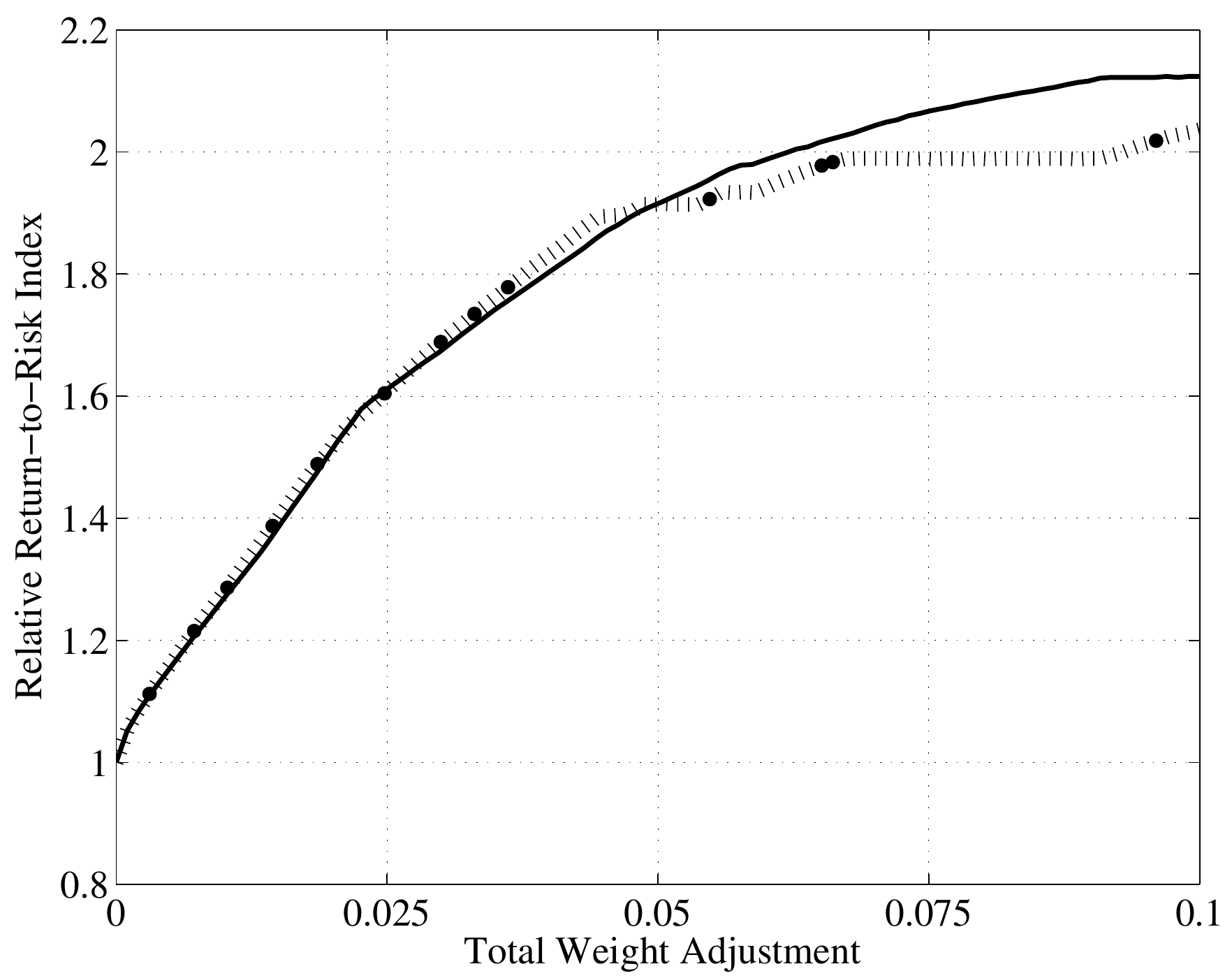}\quad
  \includegraphics[width=0.475\linewidth]{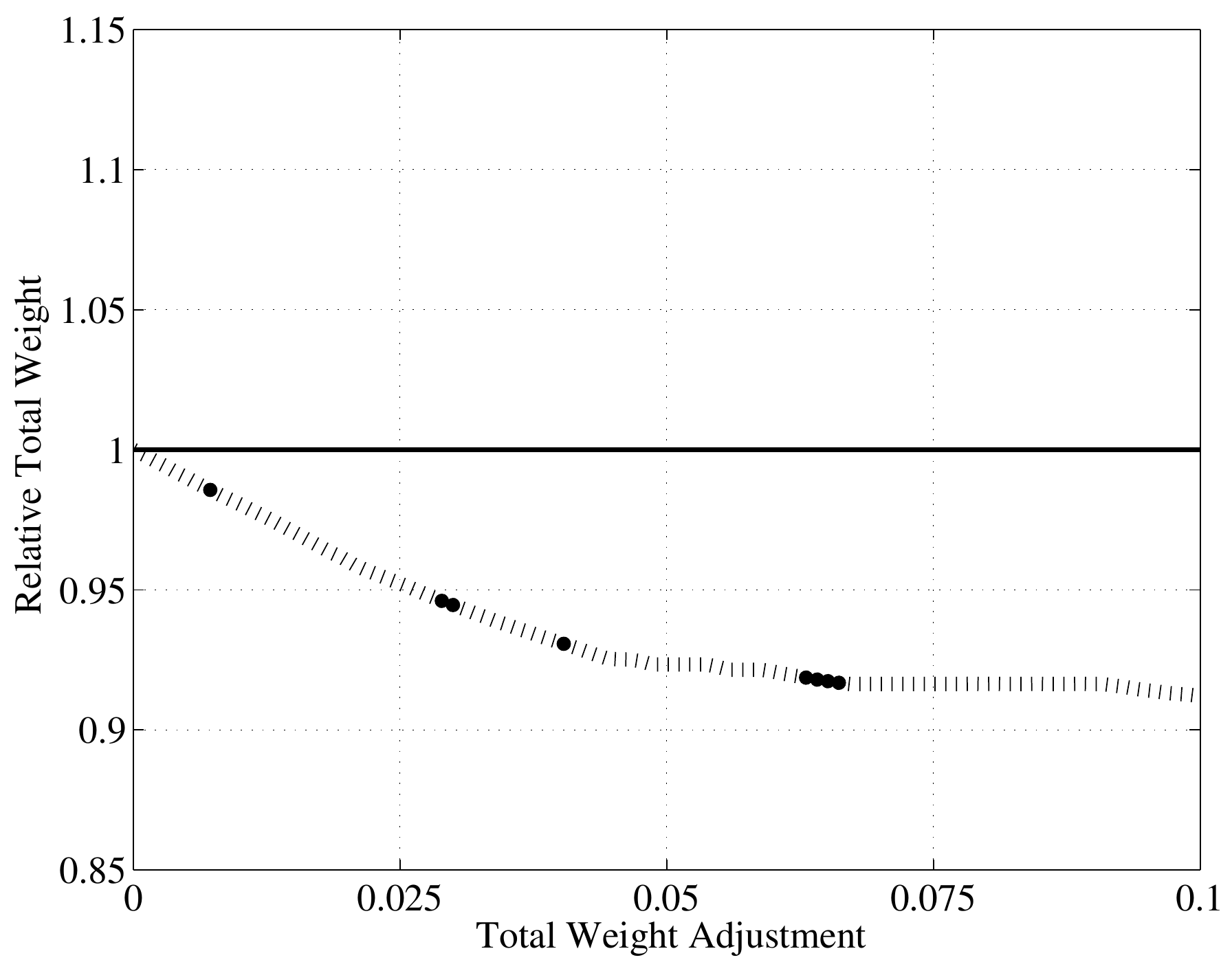}\quad\\
  \includegraphics[width=0.475\linewidth]{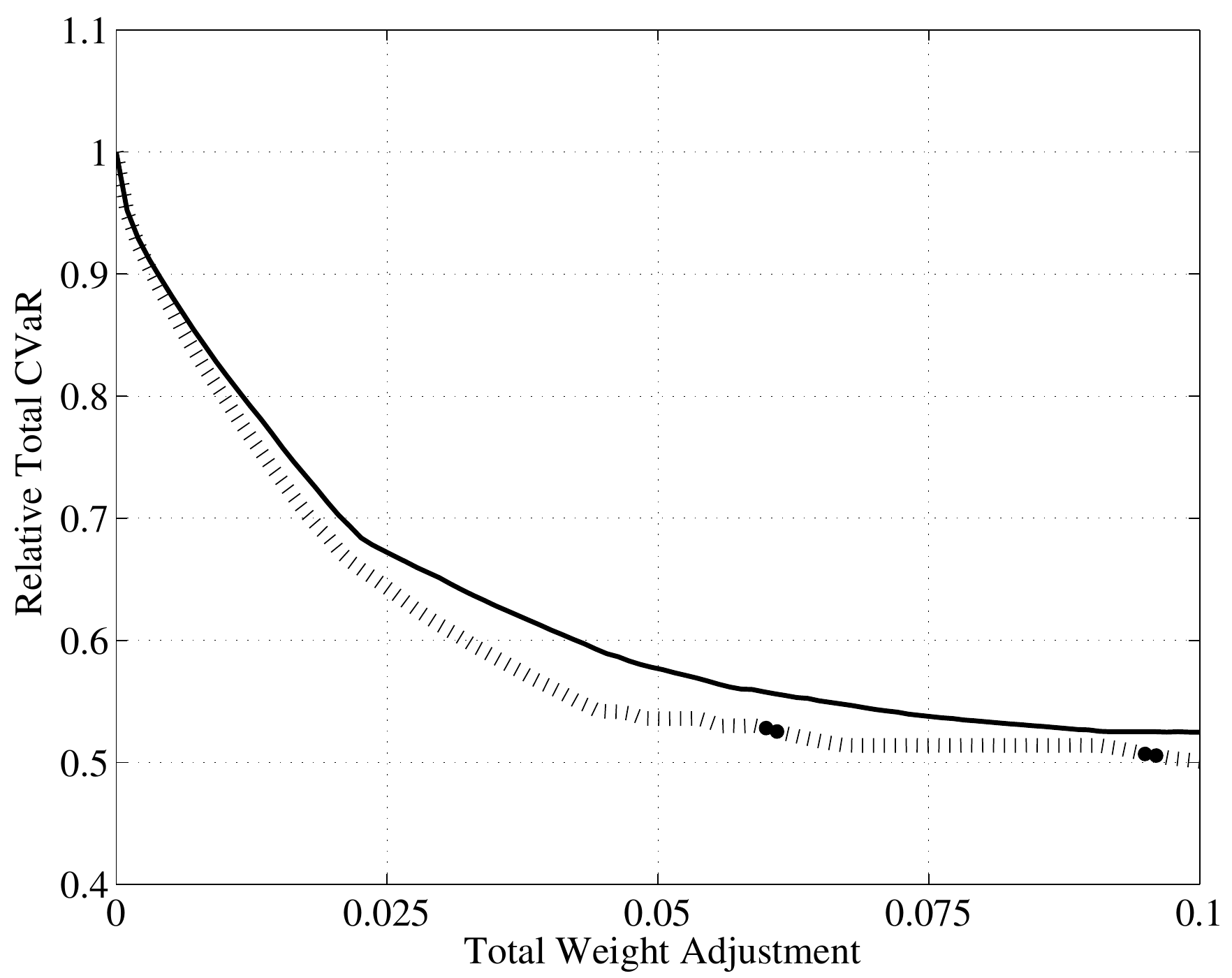}\quad
  \includegraphics[width=0.475\linewidth]{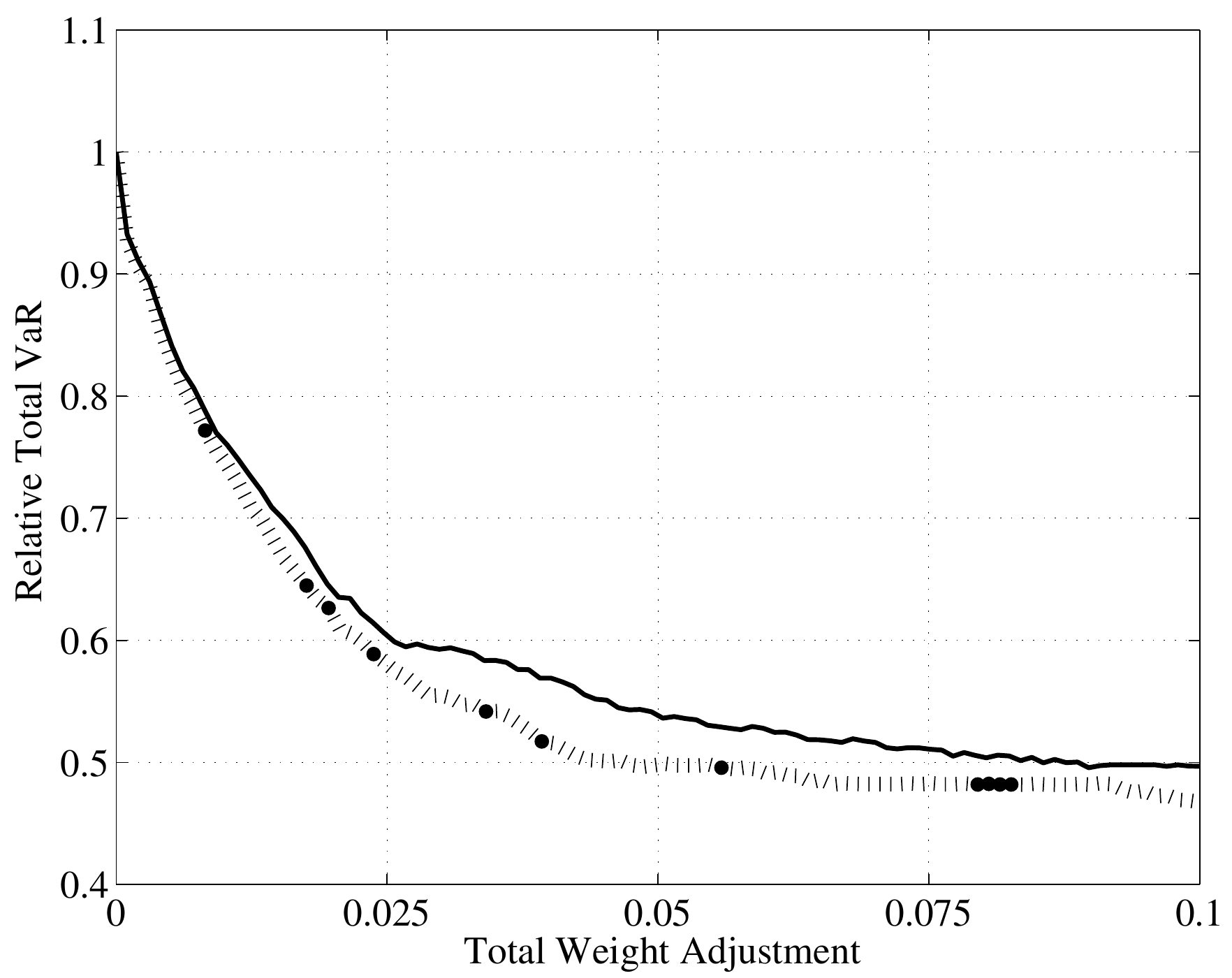}\quad\\
  \includegraphics[width=0.475\linewidth]{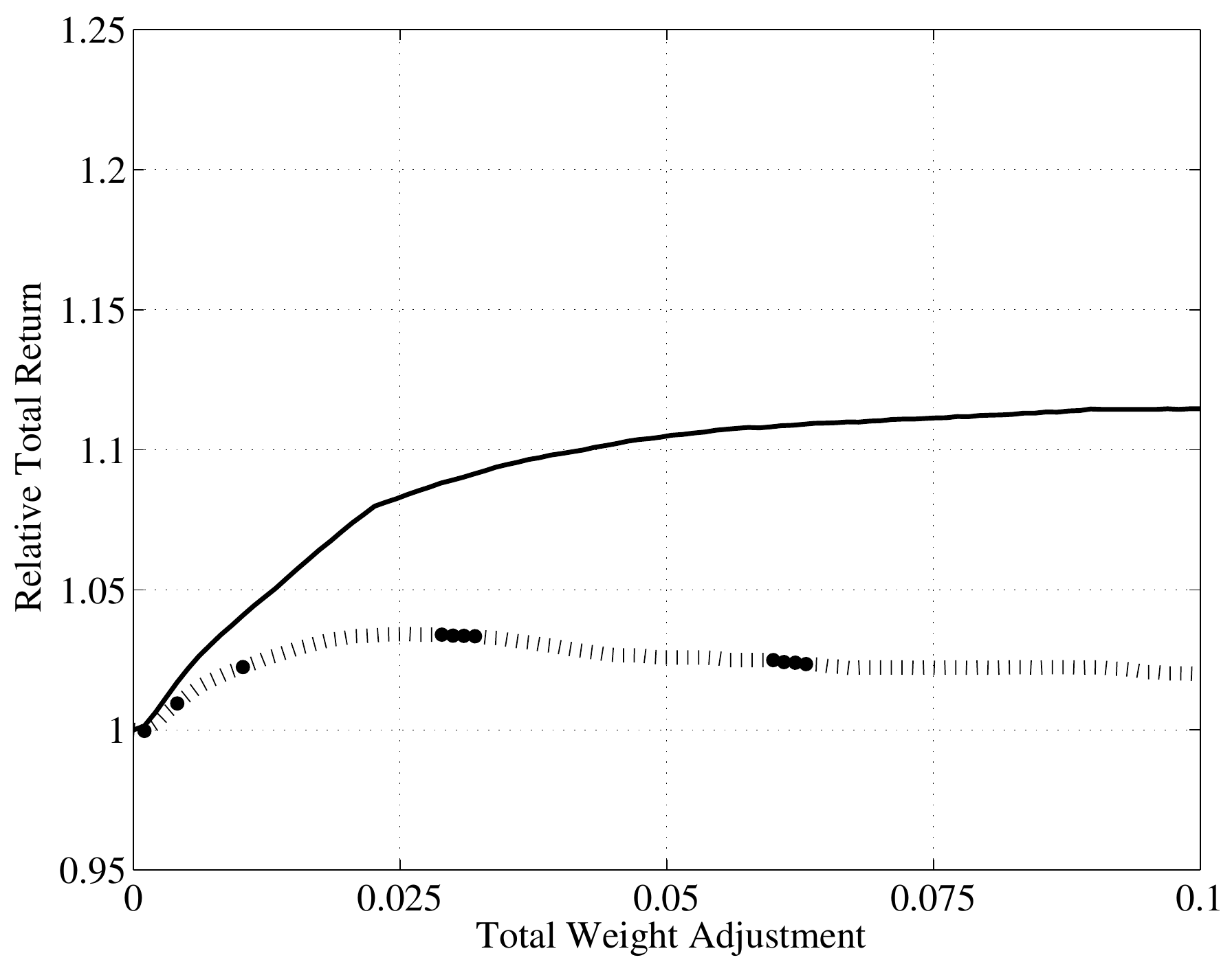}\quad
  \includegraphics[width=0.475\linewidth]{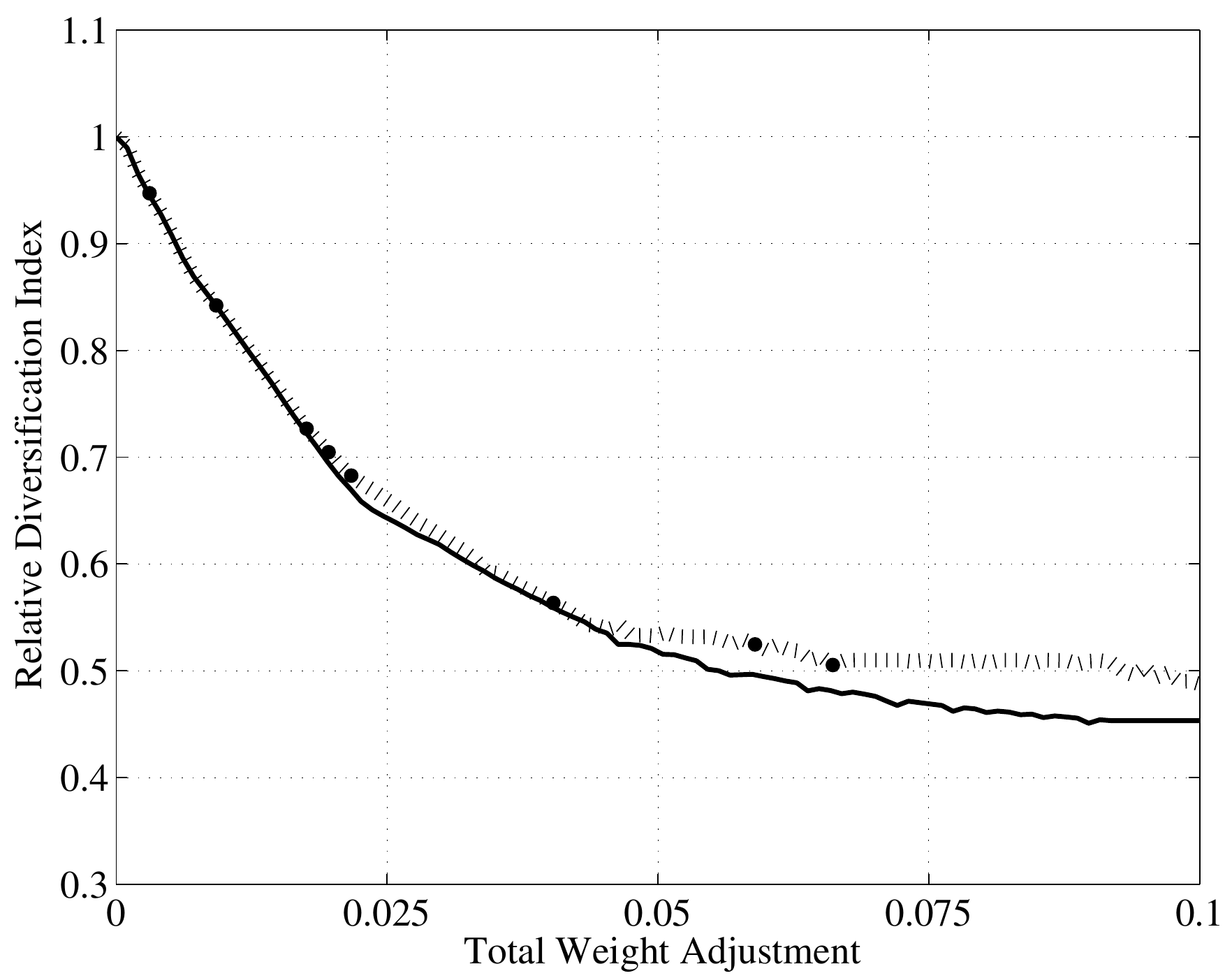}\quad\\
  \caption{\label{fig:} Comparison of return-to-risk index maximisation results relative with respect to the initial states under different constraints: fixed total revenue (---) and no constraints ($\cdots$) by varying the total amount of portfolio adjustment from 0 to 0.1 when $\delta c=10^{-5}$.}
\end{figure}

The numerical procedure is, at best, locally first-order accurate with respect to $\delta c$ (see Figure 9 and Table 2). For our particular numerical implementation of risk minimisation under the fixed total revenue when the total amount of portfolio adjustment goes from 0 to 0.01, at the latter of which the total portfolio risk has reached a steady state, the rate of convergence is approximately 0.5389, which is less than 1. This is probably owing to missing discontinuities occasionally during our numerical continuation procedure.

\begin{figure}
  \centering
    \includegraphics[width=0.7\linewidth]{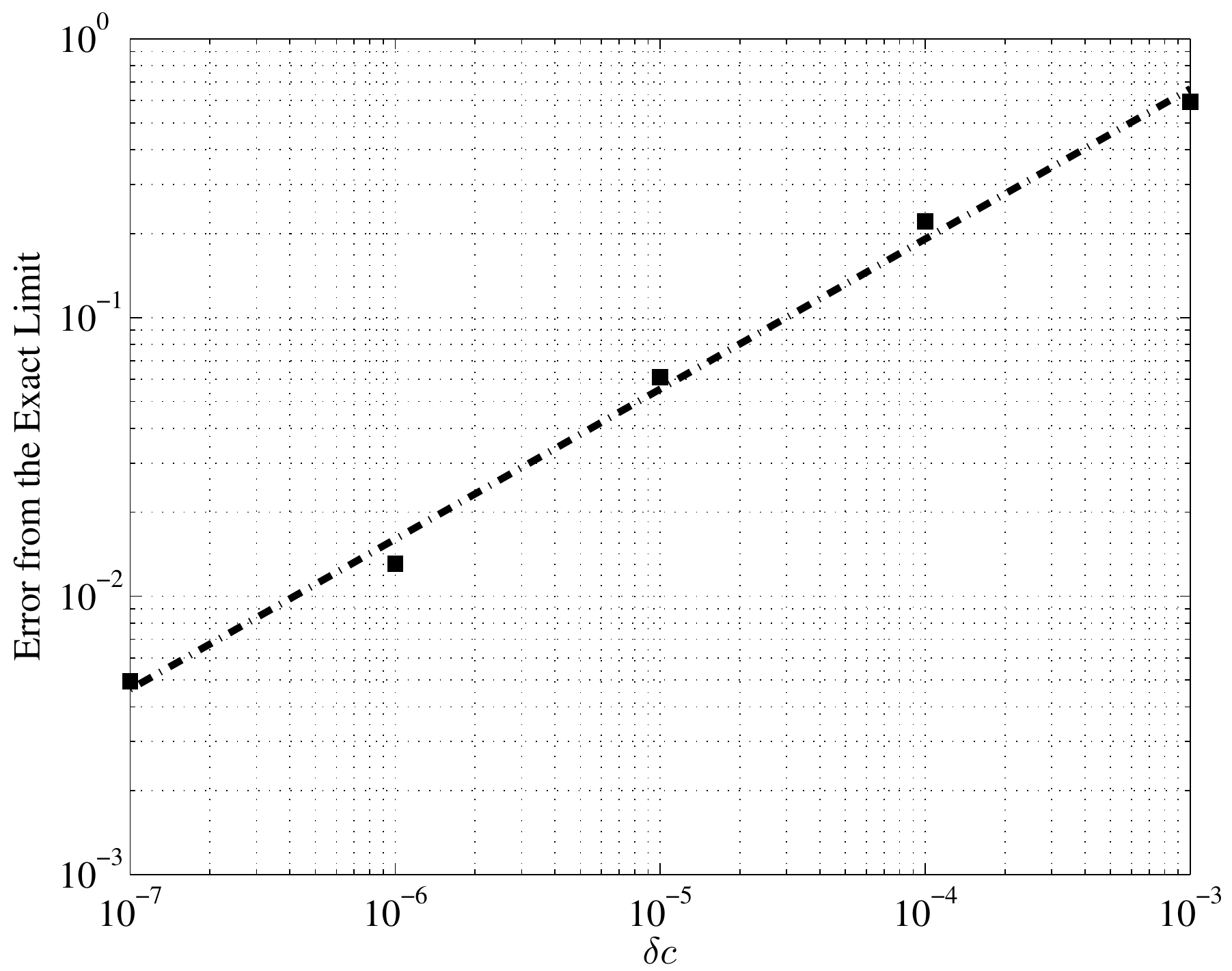}
   \caption{\label{fig:} The rate of convergence of risk minimisation under the fixed total revenue. The error from the exact limit of the steady total risk relative to the initial one at $c_{M}=0.01$ is assumed to be approximately $A\delta c^d$ where $A\approx 27.3759$ and $d\approx 0.538879$. The average logarithmic least-square residual is about 0.0175859.}
\end{figure}

\begin{table}
\tbl{The relative numerical steady-state total risks and their error from the exact limit versus $\delta c$ ranging from $10^{-7}$ to $10^{-3}$ for the total amount of portfolio adjustment being 0.01.}
{
\begin{tabular}[|]{@{}cccc}	
\toprule
$\delta c$ & $\frac{{\rm CVaR}_{\beta}^{({\rm loss})}(X_M)}{{\rm CVaR}_{\beta}^{({\rm loss})}(X_0)}$ & Error from the exact limit & Log-error from the exact limit\\
\colrule
0& 0.759604 & -- & $-\infty$ \\
$10^{-7}$ & 0.764546 & 0.00494238 & $-5.30991$ \\
$10^{-6}$ &0.772676 & 0.0130720 & $-4.33728$ \\
$10^{-5}$ &0.820603 & 0.0609996 & $-2.79689$ \\
$10^{-4}$ &0.981190 & 0.221586 & $-1.50694$ \\
$10^{-3}$ &1.35353 & 0.593922 & $-0.521007$ \\
\botrule
\end{tabular}
}
\label{table2}
\end{table}

\section{Discussion}

Some criticism may arise at the preliminary stage that general distribution functions are not used for the purpose of portfolio optimisation: the initial portfolio data at hand is just a realised or projected instance of all possible plausible scenario for the behaviour of a portfolio. However, our approach here is acceptable because the main objective is to find the best possible way to adjust the amount of each asset or asset group in a portfolio only on the basis of the realised history or a virtual scenario of the portfolio. The essentially same idea has been earlier used in \cite{RockafellarUryasev2000, Uryasev2000, KrokhmalPalmquistUryasev2002,  MansiniOgryczakSperanza2007} as well, so we do not intend that our discussion goes further to get an ultimate answer for the portfolio optimisation starting from the pre-assigned distribution for the value of the whole portfolio along with a well-defined correlation between its component assets or asset groups. Even for the latter (ultimate) case, the portfolio distribution and its correlation should be modelled in most occasions, according to empirical scenarios for portfolio performance, otherwise mentioned. Accordingly, the way of our confined discussion, {\it per se}, is meaningful.

A small neighbourhood around a certain portfolio state needed to proceed an optimisation procedure can be any type of small ball measured by $l_p$ for $p>0$ centerred at the state in the weight distribution space. When $p=1$, the Lagrange multiplier method can be translated into an ordinary LP formulation. When $p=2$, which is the case in this paper, the required analytical procedure explicitly ends up with a quadratic closed form that greatly enhances the computational cost. The size of ellipsoid should be taken small enough, in principle, to search the right optimal path even if an occasional jumping over happens between piecewise linear hyper-plane segments of different normal vectors. However, taking too much small ellipsoid may cause the computational time to be excessive if every such jumping over is tried to be avoided.

The order of computational complexity is $O(MNK\log{K})$. The total computation time for our risk minimisation test problems with $N=252$ asset groups and $K=2\times 10^3$ scenarios for $M=10^4$ iterations using Intel\textsuperscript{$\circledR$} Core\textsuperscript{TM} i5-2500 CPU @ 3.30GHz in MATLAB is about $5\times 10^3$ seconds, no matter what objective functions and constraints are used.

The path independence of the optimal solution in the parameter space is not necessarily guaranteed, in general because the optimal solution path may vary depending on how to choose the parameter path. The perfect smoothness of the optimal solution path is not necessarily guaranteed, either, when the probability distribution function is made of a discrete set of raw time series. This affects the non-smoothness of the optimal state functions, such as the return-to-risk ratio, the total portfolio risk, the total portfolio return, and the diversification index, etc., with respect to the amount of change in the total amount of portfolio adjustment.

The amount of weight adjustments for individual assets or asset groups should be understood all comparable each other in the sense that
\begin{eqnarray}
0\ll\frac{c^{(n)}}{c^{(\tilde{n})}}
\end{eqnarray}
for all pairs of $(n,\tilde{n})$ where $n,\,\tilde{n}=1,2,\,\cdots,\,N$. The amount of weight adjustment may be affected from various factors, such as the liquidity of each asset or asset group, reallocation or transaction fee, etc. In case that the amount of weight adjustment is a nontrivial function of the weight, we may need to justify that the amount of weight adjustment is slowly varied with respect to the amount of weight readjustment.

The return maximisation may not always give the optimal results in the sense of achieving the desired directions of risk states. Depending on the risk structure of portfolio, maximising the return may sometimes lead to unfavourable excessive increase of the total portfolio risk.

The cost minimisation as an objective in the optimisation procedure and the dependence of optimal states on the confidence level are worth further investigation. There are much room for in-depth studies when stochastic returns, non-static risk distributions, or options on portfolio are considered.

The quadratic nonlinear projection method introduced in this paper is applicable for portfolio optimisation with other types of risk measures, such as the drawdown risk (see Goldberg \& Mahmoud (2014)).

\section{Concluding remarks}

The optimisation framework discussed in this paper is valid for other general types of statistical information, not only for the time series of which probability distribution is skewed or fat-tailed. This method can be a useful replacement for various portfolio optimisation problems, which do not belong to the proper realm of mean--variance analysis.

\section*{Acknowledgement(s)}
This paper was initiated from a credit portfolio optimisation project conducted at the Hana Bank, South Korea from April to July in 2008. The authors wish to thank Roger H. Kim, Junsuk Her, and Prof. Stanislav Uryasev for their kind collaboration and discussion with the authors.

\appendix

\section{Solution to the Lagrange multiplier formulation for the local single-step optimisation of the first-order variational approximation}

\subsection{The case when both of the total revenue and the total risk or return constraints are included}
The system of equations for the Lagrange multiplier method framework of our optimisation method, (\ref{LagrangeMultiplierMethod_Main})--(\ref{LagrangeMultiplierMethod_Eqns}), can be written as follows:
\begin{subequations}
\begin{eqnarray}
\label{LagrangeMultiplierMethod_Appendix_A1a}
\frac{\delta \mathcal{L}}{\delta y^{(n)}}&=&f^{(n)}-s-h^{(n)}t-\frac{{c^{(n)}}^2y^{(n)}}{\sqrt{D}}q=0\quad {\rm for}\quad n=1,\,2,\,\cdots,\,N,\\
\label{LagrangeMultiplierMethod_Appendix_A1b}
\frac{\partial \mathcal{L}}{\partial s}&=&\sum\limits_{n=1}^N y^{(n)}-\kappa_1=0,\\
\label{LagrangeMultiplierMethod_Appendix_A1c}
\frac{\partial \mathcal{L}}{\partial t}&=&\sum\limits_{n=1}^{N}h^{(n)}y^{(n)}-\kappa_2=0,\\
\label{LagrangeMultiplierMethod_Appendix_A1d}
\frac{\partial \mathcal{L}}{\partial q}&=&\sqrt{D}-1=0,
\end{eqnarray}
\end{subequations}
where $D=\sum\limits_{n=1}^{N}{c^{(n)}}^2{y^{(n)}}^2$. From (\ref{LagrangeMultiplierMethod_Appendix_A1a}), $y^{(n)}$ is solved to be
\begin{eqnarray}
\label{Solution_y_n_pm}
y^{(n)}&=&\frac{f^{(n)}-s-h^{(n)}t}{q{c^{(n)}}^2}\quad {\rm for}\quad n=1,\,2,\,\cdots,\,N.
\end{eqnarray}

Substituting (\ref{Solution_y_n_pm}) into (\ref{LagrangeMultiplierMethod_Appendix_A1b})--(\ref{LagrangeMultiplierMethod_Appendix_A1c}), we have the following system of linear equations for $s$ and $t$:
\begin{subequations}
\begin{eqnarray}
\left(
\begin{array}{cc}
U & V\\
V & W
\end{array}
\right)
\left(
\begin{array}{c}
s \\ t
\end{array}
\right)
&=&
\left(
\begin{array}{c}
G-\kappa_1q \\H-\kappa_2q
\end{array}
\right),\\
\label{expression_s_Appendix}
s&=&\frac{(\kappa_2 V-\kappa_1 W)q+GW-HV}{UW-V^2},\\
\label{expression_t_Appendix}
t&=&\frac{(\kappa_1 V-\kappa_2 U)q+HU-GV}{UW-V^2},
\end{eqnarray}
\end{subequations}
from which (\ref{Solution_y_n_pm}) yields (\ref{expression_dw_dc}), where $G$, $H$, $U$, $V$, and $W$ are defined in (\ref{Constant_G})--(\ref{Constant_W}).

Finally, substituting (\ref{expression_dw_dc}) into (\ref{LagrangeMultiplierMethod_Appendix_A1d}), a quadratic equation for $q$ is obtained by
\begin{eqnarray}
a_2q^2+a_0&=&0,
\end{eqnarray}
upon beautiful simplification, where $a_0$, $a_2$, and $F$ are defined in (\ref{Constant_F}) and (\ref{expression_a0})--(\ref{expression_a2}). It turns out that the linear term in the quadratic equation for $q$ vanishes out. $a_0$ is always non-negative because it is calculated by squaring all the terms that do not involve $q$ for the expression of $y^{(n)}$ in (\ref{expression_dw_dc}).

Also, note that
\begin{subequations}
\begin{eqnarray}
UW&\ge& V^2,\\
UF&\ge& G^2,\\
WF&\ge& H^2,
\end{eqnarray}
\end{subequations}
hence,
\begin{subequations}
\begin{eqnarray}
U\kappa_2^2+W\kappa_1^2-2V\kappa_1\kappa_2 &\ge& 0,\\
G^2W+H^2U-2GHV&\ge& 0.
\end{eqnarray}
\end{subequations}

\subsection{The case when the risk or return constraint is excluded}

From \ref{LagrangeMultiplierMethod_Main}, we set $t=0$. Then, we have
\begin{eqnarray}
\label{expression_s_option2_Appendix}
s&=&\frac{G-\kappa_1q}{U},
\end{eqnarray}
which yields (\ref{expression_dw_dc_option2}) by substituting into (\ref{LagrangeMultiplierMethod_Appendix_A1b}). The quantities $a_0$ and $a_2$ in (\ref{expression_a0_option2})--(\ref{expression_a2_option2}) for $q$ come out of (\ref{LagrangeMultiplierMethod_Appendix_A1d}) by using (\ref{expression_dw_dc_option2}).

\subsection{The case when the total revenue constraint is excluded}

For this case, $s=0$, hence, we obtain
\begin{eqnarray}
\label{expression_t_option3_Appendix}
t&=&\frac{H-\kappa_2q}{W}
\end{eqnarray}
and the expression (\ref{expression_dw_dc_option3}). Likewise, as the above, we have the expressions (\ref{expression_a0_option2})--(\ref{expression_a2_option2}) for $q$.

\subsection{The case when both of the total revenue and the total risk or return constraints are excluded}

Here, $s=t=0$, from which, the desired expression (\ref{expression_dw_dc_option4}) is obtained.

\section{Extremum points of $Q$}
\subsection{The case when both of the total revenue and the total risk or return constraints are included}

\subsubsection{Non-fixed total revenue and non-fixed total return or risk}

At the extremum points of $Q\left({\kappa}_1,{\kappa}_2\right)$ from (\ref{expression_Q}), we require that
\begin{subequations}
\begin{eqnarray}
\label{dQ_dkappa_1}
\left.\frac{\partial Q}{\partial\kappa_1}\right\vert_{(\kappa_1,\kappa_2)=(\bar{\kappa}_1,\bar{\kappa}_2)}&=&-\frac{W\bar{\kappa}_1 -V\bar{\kappa}_2}{UW-V^2}\bar{q}+\frac{GW-HV}{UW-V^2}=0,\\
\label{dQ_dkappa_2}
\left.\frac{\partial Q}{\partial\kappa_2}\right\vert_{(\kappa_1,\kappa_2)=(\bar{\kappa}_1,\bar{\kappa}_2)}&=&-\frac{U\bar{\kappa}_2-V\bar{\kappa}_1}{UW-V^2}\bar{q}+\frac{HU-GV}{UW-V^2}=0
\end{eqnarray}
\end{subequations}
for
\begin{eqnarray}
\bar{q}&=&\pm\left.\sqrt{-\frac{a_0}{a_2}}\right\vert_{(\kappa_1,\kappa_2)=(\bar{\kappa}_1,\bar{\kappa}_2)},
\end{eqnarray}
where $a_0$ and $a_2$ are defined as in (\ref{expression_a0})--(\ref{expression_a2}).

The above system of equations yields a relation between $\bar{q}$ and $(\bar{\kappa}_1,\bar{\kappa}_2)$, given by
\begin{subequations}
\begin{eqnarray}
\label{extremum_points_1}
(\bar{\kappa}_1,\bar{\kappa}_2)&=&\frac{\left(G,H\right)}{\bar{q}},
\end{eqnarray}
then solved to be
\begin{eqnarray}
\label{extremum_points_2}
\bar{q}^2&=&F.
\end{eqnarray}
\end{subequations}
This pair $\left(\bar{\kappa}_1,\bar{\kappa}_2\right)$ naturally gives the steepest descent or ascent direction for the objective function $Q$. It turns out that this extremum condition is satisfied exactly when $s=t=0$.

These extreme points are exactly the maximum and the minimum of $Q$. This is because the Hessian of $Q$ at those points becomes
\begin{eqnarray}
\mathbf{H}\left[Q\right](\bar{\kappa}_1,\bar{\kappa}_2)&=&\frac{F^2}{(UW-V^2)a_0},
\end{eqnarray}
which is always positive for any non-trivial portfolio. In addition, the second-order partial derivatives of $Q$ with respect to $\kappa_1$ and $\kappa_2$ at the extremum points are obtained by
\begin{subequations}
\begin{eqnarray}
\label{d2Q_dk1_2}
\left.\frac{\partial^2 Q}{\partial\kappa_1^2}\right\vert_{(\kappa_1,\kappa_2)=(\bar{\kappa}_1,\bar{\kappa}_2)}&=&-\left\{\frac{W}{UW-V^2}+\frac{(GW-HV)^2}{(UW-V^2)^2a_0}\right\}\bar{q},\\
\label{d2Q_dk2_2}
\left.\frac{\partial^2 Q}{\partial\kappa_2^2}\right\vert_{(\kappa_1,\kappa_2)=(\bar{\kappa}_1,\bar{\kappa}_2)}&=&-\left\{\frac{U}{UW-V^2}+\frac{(HU-GV)^2}{(UW-V^2)^2a_0}\right\}\bar{q},
\end{eqnarray}
\end{subequations}
in which the signs are all flipped from $\bar{q}$.

Therefore, the global maximum and minimum of $Q$ at the local constraints are found to be $\pm\sqrt{F}$ at $\bar{q}=\pm\sqrt{F}$.

\subsubsection{Fixed total revenue: $\kappa_1\equiv 0$}

Setting $\kappa_1\equiv 0$ from (\ref{dQ_dkappa_2}), we have
\begin{subequations}
\begin{eqnarray}
\bar{q}&=&\pm\sqrt{F-\frac{G^2}{U}},\\
\bar{\kappa}_2&=&\frac{HU-GV}{U\bar{q}}
\end{eqnarray}
\end{subequations}
when $t=0$. Note that $UF-G^2>0$ for any non-trivially distributed portfolio. For this condition, $\delta\gamma$ cannot be zero. The criterion to choose the sign is the same as before.

\subsubsection{Fixed total return or risk: $\kappa_2\equiv 0$}

Setting $\kappa_2\equiv 0$ from (\ref{dQ_dkappa_1}), we have
\begin{subequations}
\begin{eqnarray}
\bar{q}&=&\pm\sqrt{F-\frac{H^2}{W}},\\
\bar{\kappa}_1&=&\frac{GW-HV}{W\bar{q}}
\end{eqnarray}
\end{subequations}
when $s=0$. For any non-trivially distributed portfolio, $WF-H^2>0$, likewise, and $\delta\alpha\ne 0$.

\subsubsection{Fixed total revenue and total risk or return: $\kappa_1=\kappa_2\equiv 0$}

$Q$ is independent of $(\kappa_1,\kappa_2)$, where $a_0$ is defined as in (\ref{expression_a0}) and $a_2=-1$: $\bar{q}=\sqrt{a_0}$.

\subsection{The case when the risk or return constraint is excluded}
At the extremum points of $Q(\kappa_1)$ from (\ref{expression_Q_option2}), it should follow that
\begin{eqnarray}
\left.\frac{\partial Q}{\partial \kappa_1}\right|_{\kappa_1=\bar{\kappa}_1}&=&-\frac{\bar{\kappa}_1}{U}\bar{q}+1=0
\end{eqnarray}
for
\begin{eqnarray}
\bar{q}&=&\pm\left.\sqrt{-\frac{a_0}{a_2}}\right|_{\kappa_1=\bar{\kappa}_1},
\end{eqnarray}
where $a_0$ and $a_2$ are given as in (\ref{expression_a0_option2})--(\ref{expression_a2_option2}).

Then, this is solved by
\begin{subequations}
\begin{eqnarray}
\bar{q}&=&\pm\sqrt{U+a_0},\\
\bar{\kappa}_1&=&\frac{U}{\bar{q}}.
\end{eqnarray}
\end{subequations}

\subsection{The case when the total revenue constraint is excluded}

Likewise, as the above case, $Q(\kappa_2)$ attains the extremum values when
\begin{subequations}
\begin{eqnarray}
\bar{q}&=&\pm\sqrt{W+a_0},\\
\bar{\kappa}_2&=&\frac{W}{\bar{q}}
\end{eqnarray}
\end{subequations}
for
\begin{eqnarray}
\bar{q}&=&\pm\left.\sqrt{-\frac{a_0}{a_2}}\right|_{\kappa_2=\bar{\kappa}_2},
\end{eqnarray}
where $a_0$ and $a_2$ are as in (\ref{expression_a0_option3})--(\ref{expression_a2_option3}).

\subsection{The case when both of the total revenue and the total risk or return constraints are excluded}

$Q$ is independent of $(\kappa_1,\kappa_2)$, so it is unnecessary to consider the extremum points of $Q$ for this case: $Q=\bar{q}=\sqrt{F}$.

\end{document}